\documentclass[pdflatex,sn-mathphys-num]{sn-jnl}%

\usepackage{colortbl}  
\usepackage[table,xcdraw]{xcolor}

\usepackage{multirow}%
\usepackage{amsmath,amssymb,amsfonts}%
\usepackage{amsthm}%
\usepackage{mathrsfs}%
\usepackage[title]{appendix}%
\usepackage{textcomp}%
\usepackage{manyfoot}%
\usepackage{booktabs}%
\usepackage{algorithm}%
\usepackage{algorithmicx}%
\usepackage{algpseudocode}%
\usepackage{listings}%
\usepackage{tabularx}
\usepackage{makecell}
\usepackage{tabularray}
\usepackage{hyperref}
\usepackage{stfloats}
\usepackage{balance}
\usepackage{url}
\usepackage{verbatim}
\usepackage{hyperref}
\usepackage[T1]{fontenc}
\usepackage{subfig}
\usepackage{graphicx}
\usepackage{xcolor}
\usepackage{tabularray}
\usepackage[table]{xcolor} 
\usepackage{graphicx}
\usepackage{tabularx} 
\usepackage{booktabs} 

\theoremstyle{thmstyleone}%
\theoremstyle{thmstyletwo}%

\theoremstyle{thmstylethree}%

\begin{document}

\title[Article Title]{Towards a Zero Trust Decentralized Identity Management System for Secure Autonomous Vehicles}

\author*[1]{\fnm{Amal} \sur{Yousseef}}\email{amalyousseef@arizona.edu}

\author[1]{\fnm{Shalaka} \sur{Satam}}\email{shalakasatam@arizona.edu}

\author[1]{\fnm{Banafsheh Saber} \sur{Latibari}}\email{banafsheh@arizona.edu}

\author[1]{\fnm{Mai} \sur{A. Abdel-Malek}}\email{mmalek@arizona.edu}

\author[1]{\fnm{Soheil} \sur{Salehi}}\email{ssalehi@arizona.edu}

\author[1,2]{\fnm{Pratik} \sur{Satam}}\email{pratiksatam@arizona.edu}

\affil[1]{\orgdiv{Electrical and Computer Engineering},\orgname{University of Arizona}, \city{Tucson}, \postcode{85719}, \state{AZ}, \country{USA}}

\affil[2]{\orgdiv{Systems and Industrial Engineering}, \orgname{University of Arizona}, \city{Tucson}, \orgaddress{\postcode{85719}, \state{AZ}, \country{USA}}}

\abstract{Autonomous vehicles (AVs) rely on pervasive connectivity to enable cooperative and safety-critical applications, but this connectivity also exposes them to a wide range of cybersecurity threats. Existing perimeter-based security and centralized identity management approaches are inadequate for highly dynamic V2X environments, as they depend on implicit trust and suffer from scalability and single-point-of-failure limitations. This paper proposes D-IM, a Zero Trust-based decentralized identity management and authentication framework for secure V2X communication. D-IM integrates continuous verification with a permissioned blockchain to eliminate centralized trust assumptions and enforce explicit, verifiable identity relationships among vehicles and infrastructure. The framework is designed around clear Zero Trust-aligned goals, including mutual authentication, decentralization, privacy protection, non-repudiation, and traceability, and addresses a comprehensive attacker model covering identity, data integrity, collusion, availability, and accountability threats. We present the D-IM system architecture and identification and authorization protocol, and validate its security properties through both qualitative analysis and a formal BAN logic-based verification. Simulation results in urban and highway scenarios using DSRC and C-V2X demonstrate that D-IM introduces limited overhead while preserving network performance, supporting its practicality for real-world AV deployments.}




\maketitle

\section{INTRODUCTION}
Autonomous Vehicles (AVs) are revolutionizing modern transportation by offering improved road safety, efficiency, and convenience, while minimizing human error that is responsible for more than 90\% of road accidents and more than 40,000 fatalities and 2 million injuries annually in the United States alone \cite{singh2015critical,national2020early}. AVs integrate sophisticated sensors, Electronic Control Units (ECUs), and advanced communication protocols to enable semi- and fully autonomous driving. These systems rely on Global Positioning Systems (GPS), LiDAR, cameras, and Vehicle-to-Everything (V2X) communication for perception, situational awareness, and decision-making \cite{bhatia2019autonomous}.  

The Society of Automotive Engineers (SAE) defines six levels of driving automation, ranging from Level~0 (no automation) to Level~5 (full automation). These levels describe the degree of vehicle autonomy in terms of steering control, environment monitoring, and fallback responsibility \cite{satam2022autonomous,yousseef2025autonomous}. Table~\ref{tab:level_sae} summarizes the SAE autonomy levels. As AVs progress toward higher autonomy (aiming for Level 5), their reliance on connectivity becomes indispensable. V2X communication, including Vehicle-to-Vehicle (V2V), Vehicle-to-Infrastructure (V2I), and Vehicle-to-Network (V2N) enables cooperative mobility. Key use cases include platooning, where vehicles form convoys to reduce drag and improve fuel efficiency; drone swarms, coordinating with AVs for delivery or traffic monitoring; emergency vehicle prioritization, where AVs yield dynamically to ambulances and police; and cooperative perception, where vehicles share sensor data for beyond-line-of-sight awareness \cite{ying2024literature}. Dedicated Short-Range Communications (DSRC) and cellular-V2X (C-V2X) are the primary enabling radio access technologies, supporting low-latency and high-reliability communication.

\begin{table}[h]
  \caption{SAE Vehicle Autonomy levels}
  \begin{tabular}{|c|c|c|c|c|c|}
    \hline
    \rowcolor[HTML]{EFEFEF} \textbf{Level} & \textbf{Automation} & \textbf{\begin{tabular}[c]{@{}c@{}}Steering \\ Cruising\end{tabular}} & \textbf{\begin{tabular}[c]{@{}c@{}}Environment \\ Monitoring\end{tabular}} & \textbf{\begin{tabular}[c]{@{}c@{}}Fallback \\ Control\end{tabular}} & \textbf{\begin{tabular}[c]{@{}c@{}}Driving \\ Mode\end{tabular}} \\ \hline
    0 & None        & H & H & H & N/A \\ \hline
    1 & Supportive  & H,S & H & H & Some \\ \hline
    2 & Partial     & S & H & H & Some \\ \hline
    3 & Conditional & S & S & H & Some \\ \hline
    4 & High        & S & S & H & Some \\ \hline
    5 & Full        & S & S & S & All \\ \hline
  \end{tabular}%
  \label{tab:level_sae}
\end{table}



In platooning, for instance, the lead vehicle periodically broadcasts kinematic states (e.g., position, speed, acceleration) and intent, while the follower vehicles fuse these messages with onboard sensors to compute control inputs that maintain inter-vehicle gaps and string stability where the disturbances are not amplified when propagating along the vehicles string. While this cooperative driving strategy improves safety and fuel efficiency, it also introduces significant vulnerabilities. The increased interconnection introduces severe cybersecurity concerns. For instance, malicious actors can impersonate a leader vehicle, injecting false acceleration or braking commands that endanger the convoy.  Once inside, the adversary could launch attacks such as fake data injection, transmitting falsified distance or braking information that destabilizes the convoy, or GPS spoofing, misleading the platoon about its true location and causing unsafe maneuvers. These manipulations not only threaten the cohesion of the platoon but also heighten the risk of chain-reaction accidents. Such scenarios highlight the necessity of robust identity verification and trust management mechanisms to safeguard cooperative vehicular systems.

The increased connectivity and complexity of AVs increases AVs cybersecurity vulnerabilities introducing many attack vectors. Attack vectors such as GPS spoofing, sensor manipulation, V2X spoofing and information injection threaten vehicle safety and operational integrity. Real-world incidents, such as the Jeep Cherokee and Tesla Model~S exploits, have demonstrated the potential for remote hijacking and malicious manipulation of vehicle systems \cite{miller2015remote}. These incidents highlight the inadequacy of traditional perimeter-based security models, which assume trust within a protected network, and the urgent need for robust multilayered defense strategies \cite{anderson2023zero}. Unfortunately, many traditional approaches rely on trusted entities for credential management, a critical weakness adversaries can exploit by stealing or maliciously issuing credentials.

Identity management thus becomes critical for validating trust in dynamic vehicular environments. Current solutions primarily rely on Public Key Infrastructure (PKI) and pseudonym certificates, but these approaches face significant challenges in credential revocation, often failing to achieve timely and efficient invalidation of compromised credentials, thereby exposing vehicles to prolonged risk \cite{ndss2024Revocation}. Recent research highlights that secure and scalable identity management remains one of the most pressing V2X research challenges \cite{ying2024literature,yousseef2025autonomous}.




To secure AVs against the previously mentioned attacks, Zero Trust Architecture (ZTA) addresses limitations of perimeter models by eliminating implicit trust and enforcing continuous verification of all entities, whether inside or outside the network. ZTA aligns with AV needs where real-time decisions depend on authenticated, accurate data, and is advocated by NIST for modern, interconnected systems \cite{nist800207}. Within this paradigm, identity management emerges as the core security enabler for V2X. Ensuring that only authenticated, authorized, and non-revoked entities can publish or act upon safety messages is essential, especially in dense, dynamic traffic. Existing Identity management approaches suffer from centralization, scalability bottlenecks, and slow or bandwidth-heavy revocation mechanisms where a single trusted authority with implicit trust is needed presention a potential single point of failure \cite{ying2024v2xReview,ndss2024Revocation}. 


This paper presents a Zero Trust-based Decentralized Identity Management (D-IM) system for AVs, instantiated on a permissioned blockchain.  Our solution eliminates dependence on centralized authorities, supports (through the use of blockchain technology) low-latency identity proofing that meets V2X timing budgets, enables fast and tamper-evident revocation propagation across dynamic topologies, and scales to dense traffic scenarios while preserving privacy through unlinkable pseudonym use. Moreover, D-IM leverages blockchain-backed identifiers and verifiable credentials to bind cryptographic identities to vehicles and infrastructure, enforces continuous verification at enrollment and message time, and supports fast, tamper-evident revocation. \cite{yousseef2025autonomous,nist800207}  Conceptually, every entity, vehicle, sensor, or infrastructure node, must prove authenticity before joining the network. Identity proofs accompany safety messages so consumers can validate origin, freshness, and authorization before actuation. The decentralized ledger provides immutable auditability and eliminates implicit trust in perimeter boundaries, while Zero Trust policies restrict privileges to the minimum necessary and re-evaluate them continuously. Practically, our design leverages Hyperledger Iroha to enable lightweight decentralized consensus for ID verification, thereby mitigating risks like Sybil attacks in platooning. A malicious entity attempting to impersonate a platoon leader would be immediately detected, isolated, and denied participation, ensuring safety and integrity. Collectively, the design addresses CA centralization, revocation latency, and linkability risks identified in prior work \cite{he2022survey,miglani2023blockchain,ying2024v2xReview,ndss2024Revocation, yousseef2025autonomous, vangala2020blockchain}. 

We conduct both a qualitative and a formal security analyses using Burrows-Abadi-Needham (BAN) logic to demonstrate that the proposed framework effectively addresses the attacks defined in the threat model. We validate our framework through both conceptual analysis and preliminary simulations. The simulations cover both urban and highway scenarios for two separate radio access technologies, namely Dedicated Short Range Communications (DSRC) and  Cellular Vehicle-to-Everything (C-V2X). These simulations show minimal impact on network performance while ensuring robust identity management. Our evaluations in urban and highway scenarios using C-V2X technology indicate practical overheads: less than 19\% reduction in Packet Reception Ratio (PRR) in urban settings, 3\% in highway settings (200~vehicles/km), under 16\% increase in Channel Busy Ratio (CBR), and under 0.032~s added latency. Our framework's authentication introduces negligible delay and only minor overhead in communication metrics, demonstrating practicality for real-world deployment

\noindent Our contributions can be summarized as follows:
\begin{itemize}
    \item{We propose Zero Trust-based Blockchain-enabled ID Management and Authentication for AVs.} 
    \item{We present a qualitative and formal BAN logic-based security analysis demonstrating robustness against the considered threat model.}
    \item{ We perform a realistic urban/highway simulations reflecting dense, dynamic vehicular conditions and cooperative applications}
\end{itemize}

The remainder of this paper is structured as follows. Section~\ref{sec:relatedWork} reviews related work. Section~\ref{sec:dim_model} presents the D-IM design goals, system and attacker models.  Section~\ref{sec:dim_design} details the proposed D-IM system architecture and protocol. Section~\ref{sec:sec_analysis} provides a formal and qualitative security analysis. Section~\ref{sec:sysEval} presents performance evaluations and comparisons. Section~\ref{sec:cons} concludes and outlines future directions.

\section{RELATED WORK} \label{sec:relatedWork}
Research on securing autonomous vehicles and connected transportation systems spans multiple domains, from connected vehicle applications and cybersecurity threats to identity management. A broad body of work has examined the vulnerabilities of increased connectivity, and the limitations of existing identity management frameworks. More recently, Zero Trust Architectures (ZTA) have been proposed as a paradigm shift from perimeter-based defenses, while blockchain has been investigated for its potential to provide decentralized trust, auditability, and tamper-resistance in vehicular networks. In this section, we review related work across these themes, highlighting key advances and limitations in: (1) AV cybersecurity threats and challenges, (2) Zero Trust principles and evolution, (3) Zero Trust and blockchain integration, and (4) blockchain in vehicular and ITS security. 

\subsection{AV Cybersecurity Threats and Challenges}
Despite the transformative potential of connected and autonomous vehicles, they expose systems to a broad and evolving set of cybersecurity threats. Early demonstrations, such as the Jeep Cherokee and Tesla Model~S exploits, highlighted how attackers could remotely manipulate safety-critical systems, including braking and steering, through vulnerabilities in infotainment and controller area network (CAN) bus interfaces \cite{miller2015remote}. These incidents demonstrated the insufficiency of traditional perimeter-based defenses and established cybersecurity as a foundational requirement for safe AV deployment \cite{anderson2023zero,yousseef2025autonomous}.

AVs face diverse attack surfaces spanning sensors, in-vehicle networks, wireless communication links, and machine learning subsystems. Early works exploited weaknesses in a vehicle's CAN bus \cite{hoppe2008security} and demonstrated how attackers could compromise a 2014 Jeep Cherokee by reprogramming a gateway chip to access critical subsystems \cite{miller2014survey,miller2019lessons}. Similar vulnerabilities have been found in other vehicles, such as the Toyota Prius and Ford Escape, by targeting Electronic Control Units (ECUs) and head units \cite{checkoway2011comprehensive,miller2015remote}. Beyond the CAN bus, AVs are susceptible to attacks on navigation systems, such as GPS receivers, which can lead to misrouting or vehicle hijacking \cite{psiaki2016attackers,abrar2024gps}.

At the sensor layer, spoofing and jamming pose major threats. GPS signals can be spoofed/jammed, leading to misrouting or vehicle immobilization. LiDAR, radar, and camera sensors are vulnerable to occlusion, reflection, or adversarial perturbations. Recent studies demonstrate physical-world spoofing and removal attacks on LiDAR-based perception systems \cite{cao2023you, sato2025realism}, while surveys further review sensor failures and spoofing/jamming risks across AV perception stacks \cite{matos2024survey}. At the in-vehicle network layer, the CAN bus remains widely deployed, but vulnerable to injection, replay, and denial-of-service (DOS) attacks due to its lack of built-in authentication \cite{al2024can, lampe2024can}. At the wireless communication layer, V2X links can be disrupted by intelligent jamming and DOS attacks \cite{twardokus2023toward,arif2024clustered}, which degrade cooperative applications like platooning and cooperative perception. Furthermore, the integration of machine learning into perception and intrusion detection systems introduces new adversarial risks, where carefully crafted perturbations can mislead models for traffic sign recognition, lane detection, or anomaly detection \cite{chahe2023dynamic, pavlitska2023adversarial,kim2024survey}.

Survey efforts provide valuable taxonomies of these threats. Hussain and Zeadally present an early comprehensive survey of AV security challenges \cite{hussain2018autonomous}, while Amoozadeh et al. highlight vulnerabilities in cooperative driving streams \cite{amoozadeh2015security}, and Luo and Hou discuss countermeasures for intelligent and connected vehicles \cite{luo2019cyberattacks}. More recent surveys offer updated perspectives, including Sedar et al. on V2X cybersecurity mechanisms \cite{sedar2023comprehensive}, Kifor et al. on automotive cybersecurity frameworks and monitoring \cite{kifor2024automotive}, and Durlik and Dziubecki on readiness for AV cybersecurity \cite{durlik2024cybersecurity}. 
Collectively, these works confirm that AVs face a multi-layered threat landscape that continues to evolve, underscoring the need for multi-layered defenses that integrate anomaly detection, resilient communication protocols, and secure, continuous, and decentralized security frameworks. Identity management emerges as a cornerstone for securing V2X communications, ensuring that only authenticated and verified entities can participate in safety-critical data exchange.

\begin{figure*}[ht]
    \centering
    \includegraphics[width=0.97\columnwidth]{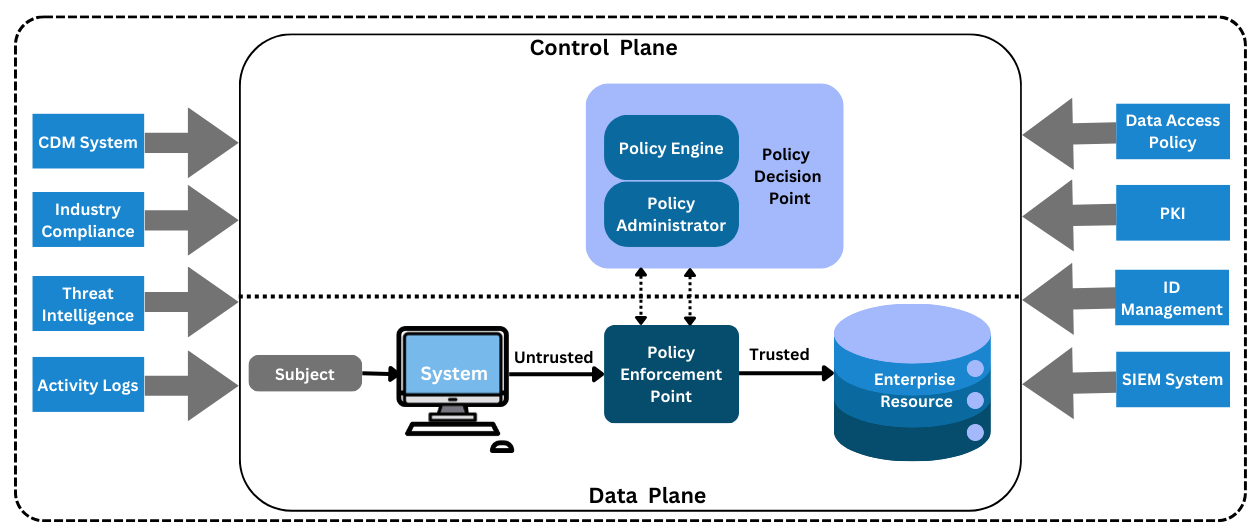}
    \caption{Zero Trust Architecture components, adapted from NIST SP 800-207.}
    \label{fig:zta_components}
\end{figure*}

\subsection{Zero Trust Principles and Evolution}
Identity management is a cornerstone of secure V2X communications, ensuring that only authenticated entities can participate in the exchange of safety-critical messages. Formal frameworks have been established by standardization bodies like IEEE STD 1609.2 \cite{7426684} and ETSI TS 103 097 \cite{etsi:103097}, which form the basis of Public Key Infrastructure ,PKI, based vehicular credential management systems. In this approach, vehicles obtain long-term certificates from a root authority and use short-term pseudonym certificates to preserve privacy and unlinkability \cite{petit2014pseudonym}. However, this anonymity comes at a cost, as regulatory and forensic requirements often necessitate conditional traceability, creating a tradeoff between privacy and accountability \cite{khodaei2015key,petit2014pseudonym,yoshizawa2023survey}.

Scalability and privacy are also ongoing limitations. High pseudonym change rates strain storage and distribution systems, while insufficient changes increase linkability risks \cite{yoshizawa2023survey}. Recent surveys emphasize that current PKI-based identity frameworks struggle to balance these needs \cite{ying2024literature,yoshizawa2023survey,almarshoud2024security,farsimadan2025review,ying2024literature,kifor2024automotive}. These limitations highlight the need for more robust, decentralized approaches to identity management, motivating the exploration of blockchain and Zero Trust principles as complementary paradigms.

Traditional perimeter-based security models, which assume trust within a protected network, are no longer sufficient for modern, interconnected systems like autonomous vehicles. To address this, the National Institute of Standards and Technology (NIST) formalized the concept of Zero Trust Architecture (ZTA) in its Special Publication 800-207, establishing the guiding principle of "never trust, always verify" \cite{nist800207}. Under this model, every access request, regardless of origin, must undergo continuous authentication and authorization. This paradigm shift has been adopted by critical sectors, including the U.S. Department of Defense \cite{dod2022zt}, and is particularly relevant for AVs, which operate in highly dynamic environments where static trust assumptions are impractical \cite{wylde2021zero, anderson2023zero, yousseef2025autonomous}.

The ZTA model comprises several key components, typically organized into a control plane and a data plane. The Policy Decision Point (PDP) consists of a Policy Engine and Policy Administrator, responsible for evaluating trust signals and determining access. The Policy Enforcement Point (PEP) enforces these decisions, mediating between untrusted and trusted zones, while data sources such as PKI, identity management, activity logs, and threat intelligence inform the policy process. Supporting systems include Continuous Diagnostics and Mitigation (CDM) systems, Security Information and Event Management (SIEM) solutions, PKI infrastructures, and identity management services. Figure~\ref{fig:zta_components} illustrates the functional decomposition of ZTA into its primary elements.

Despite its advantages, ZTA faces implementation challenges when applied to vehicular and IoT systems. Annabi et al. \cite{annabi2025survey} provide a comprehensive review of ZTA in connected vehicles, identifying open challenges in trust establishment, latency, and integration with distributed technologies. He et al. survey the state of ZTA, noting barriers such as system complexity, skill gaps, and interoperability issues, while also identifying blockchain as a potential enabler \cite{he2022survey}.  Li et al. extend this perspective to the Future Internet of Things (FIoT), arguing that ZTA can mitigate risks in massive device deployments combined with continuous verification and blockchain-assisted authentication \cite{li2022future}. The core principles of continuous verification and least privilege, however, align directly with the needs of AV ecosystems, making ZTA a foundational 
model that requires integration with decentralized trust anchors to overcome its inherent limitations \cite{yousseef2025autonomous}.

\subsection{Zero Trust and Blockchain Integration}
While ZTA provides a strong conceptual foundation through continuous verification and least privilege, its centralized decision-making components face challenges in highly dynamic environments. Blockchain emerges as a natural complement to ZTA, addressing weaknesses related to central trust, credential revocation, and tamper-proof logging \cite{dhar2021securing, alevizos2022augmenting, liu2022blockchain}. Its decentralized and immutable nature enables auditable access decisions and distributed identity verification, eliminating single points of failure. This synergy is particularly potent in the context of decentralized identity (DID) and verifiable credentials (VCs). DID frameworks allow entities to establish and prove identity without relying on centralized certificate authorities, while VCs enable selective disclosure and privacy-preserving proofs of authorization. These approaches align naturally with ZTA’s principle of continuous verification, where every request is evaluated independently of network location or prior trust \cite{vcuvcko2021decentralized, dib2020decentralized, stockburger2021blockchain}.

The integration of these paradigms has been explored across IoT and edge computing contexts, which share similar constraints with vehicular networks. For example, Dhar and Bose \cite{dhar2021securing} highlight blockchain’s role in securing IoT devices under ZTA, and Li et al. \cite{li2022zero} propose a blockchain-based zero-trust scheme for edge environments, addressing lightweight identity verification. n next-generation networks, Maksymyuk et al. \cite{maksymyuk2020blockchain} envision a blockchain-empowered framework for decentralized management in 6G, while Hassan et al. \cite{hassan2023blockchain} apply blockchain-enabled Zero Trust identity management to smart cities and IoT networks. These frameworks provide a blueprint for applying blockchain-enhanced ZT in distributed and resource-constrained ecosystems. In the vehicular domain, numerous frameworks have been proposed for blockchain-based identity management, authentication, and decentralized trust in Internet of Vehicles (IoV) and V2X \cite{george2020secure, noh2020distributed, tan2019secure, das2023secure, theodouli2020towards, agudo2020blockchain, van2017blackchain}. For instance, George et al. proposed a secure identity management framework for VANETs, while Noh et al. developed a distributed message authentication scheme.

Despite these advances, performance and scalability remain significant concerns. Blockchain protocols often introduce overhead that is ill-suited to the high mobility and low-latency demands of vehicular networks. This is due to the computational cost of cryptographic operations, the communication overhead of broadcasting transactions, and the latency of consensus protocols. Researchers have therefore focused on lightweight mechanisms for revocation and verification to balance security with real-time performance. Adja et al. \cite{adja2021blockchain} advanced certificate revocation using blockchain, while Liu et al. \cite{liu2022blockchain} proposed a decentralized, fair, and authenticated information-sharing scheme for the Internet of Things. These efforts, along with works on decentralized network management in 6G \cite{maksymyuk2020blockchain}, illustrate the ongoing challenge of reconciling blockchain's benefits with the stringent performance requirements of vehicular networks. In summary, while ZTA provides the principles and blockchain offers decentralized trust, their combined use requires optimization to address the scalability and latency challenges of current vehicular networks, motivating the framework presented in this paper.

\subsection{Blockchain in Vehicular Networks and its Security}
Blockchain has emerged as a promising enabler for Intelligent Transportation Systems (ITS), offering decentralized trust, tamper-resistant logging, and auditable data exchange. Early visionary work by Yuan and Wang \cite{yuan2016towards} highlighted the potential of blockchain to address trust and coordination challenges in ITS. Subsequent surveys and frameworks confirmed these prospects, exploring applications ranging from secure communication to resource management \cite{cocirlea2020blockchain,mollah2020blockchain,jabbar2022blockchain,das2023blockchain}. Collectively, these studies frame blockchain as a foundational technology for achieving decentralized trust in connected transportation ecosystems.

In the vehicular networking domain, blockchain has been proposed as a decentralized trust anchor for secure message exchange and system integrity. Distributed trust management schemes demonstrate how blockchain can mitigate single points of failure and enhance transparency in VANET operations \cite{inedjaren2021blockchain}. Beyond fundamental networking, blockchain has supported a wide range of cooperative vehicular applications. For instance, Li et al. \cite{li2021blockchain} proposed a blockchain-based cooperative perception framework in the IoV, enabling vehicles to securely share sensory information. Song et al. \cite{song2020blockchain} developed a blockchain-enabled cooperative positioning system that integrates deep learning to enhance location accuracy. In the forensic domain, Cebe et al. \cite{cebe2018block4forensic} introduced Block4Forensic, a lightweight blockchain framework for accident investigation, while Zhu et al. \cite{zhu2024blockchain} presented a blockchain-based accident forensics system for smart connected vehicles.

Authentication and access control have been a central focus of blockchain-based vehicular security. Tan and Chung \cite{tan2019secure} proposed a secure blockchain-assisted key management scheme for VANETs, while Noh et al. \cite{noh2020distributed} designed a distributed blockchain-based message authentication protocol. Adja et al. \cite{adja2021blockchain} advanced certificate revocation and status verification using blockchain, and Das et al. \cite{das2023secure} developed a blockchain vehicle identity management framework for ITS.

Blockchain has also been leveraged to ensure data integrity and trustworthy event reporting. Dwivedi et al. \cite{dwivedi2021blockchain} proposed an IPFS-integrated blockchain protocol for secure event storage and sharing in VANETs, while Vangala et al. \cite{vangala2020blockchain} introduced the BCAS-VADN system to authenticate accident detection and notification reports on the blockchain. Resource sharing and content distribution also benefit from blockchain. Miglani and Kumar \cite{miglani2023blockchain} developed a blockchain-based framework for content-centric vehicle-to-grid (V2G) networks, ensuring secure and efficient content dissemination. Theodouli et al. \cite{theodouli2020towards} extended blockchain to identity and trust management in IoV ecosystems, while Stockburger et al. \cite{stockburger2021blockchain} explored decentralized identity for public transportation.



Scalability and performance remain persistent challenges in blockchain-enabled vehicular security systems, as consensus mechanisms can introduce significant computational and communication overhead that constrains their applicability in highly mobile V2X environments. To mitigate these limitations, Dwivedi et al. \cite{dwivedi2023design} proposed batch authentication techniques to reduce verification overhead, while Adja et al. \cite{adja2021blockchain} and Liu et al. \cite{liu2022blockchain} introduced lightweight revocation and information-sharing mechanisms to improve efficiency. Collectively, these efforts highlight ongoing attempts to reconcile the security benefits of blockchain with the stringent performance requirements of vehicular networks.

Although substantial progress has been made in identity management, Zero Trust architectures, and blockchain-enabled ITS security, an integrated solution that combines Zero Trust principles with blockchain-based decentralized identity management specifically tailored to autonomous vehicles is still lacking. Existing approaches do not simultaneously achieve scalability and privacy-preserving authentication under the tight latency constraints imposed by dynamic vehicular networks and real-time safety-critical applications such as platooning and cooperative perception.


\section{DECENTRALIZED IDENTITY MANAGEMENT (D-IM) SYSTEM Model} \label{sec:dim_model}

Autonomous vehicles operating in Vehicle-to-Everything, V2X, environments, where vehicles and roadside units (RSUs) must authenticate each other in real time before exchanging data, must continuously exchange safety-critical information such as location, speed, and intent. Guaranteeing the authenticity of these messages requires robust identity management. Traditional PKI- and pseudonym-based systems face scalability, and centralization issues, as discussed earlier in Section~\ref{sec:relatedWork}. To address these challenges, we propose a decentralized identity management (D-IM) system that integrates blockchain as a distributed trust anchor with Zero Trust principles of continuous verification and least-privilege access. Our proposed system eliminates reliance on centralized authorities, ensures continuous verification, and leverages blockchain for immutable identity records. This section presents the overall system model, attacker model, the D-IM communication framework, and the blockchain-based identification and authorization protocol.

\subsection{Design Goals}\label{sec:design_goals}
The design goals of our approach are as follows:
\begin{itemize}
    \item \textbf{Mutual Authentication:} In a zero trust environment, to ensure the identities of sharing parties, we verify their authenticity through blockchain technology after authenication of each contributing party. This ensures that only legitimate participants are allowed to share information.
    \item \textbf{Autonomy:} The authentication and authorization process, consensus mechanism and information sharing process are governed by blockchain technology, eliminating the need for any third-party authority or centralized entity.
    \item \textbf{Anonymity:} Participants may be concerned about their personal information (such as geographical positions and other sensitive or identifying details) being exposed during the information-sharing process. Therefore, blockchain is utilized to enable the sharing of real-time traffic information without revealing the participant's true identity since a pseudo-identity is given to each vehicle after authentication success. Only the blockchain distributed ledger stores the real identity information.
    \item \textbf{Privacy Protection:} The shared information should not compromise the privacy of the authenticated vehicles, i.e. disclosing their location, to unauthenticated vehicles. To ensure this, the information is always encrypted.
    \item \textbf{Non-Repudiation and Traceability of Origin:} The IDs and keys used will be stored on the blockchain distributed ledger to trace the history of information sharing.
\end{itemize}

\subsection{System Model} \label{sec:system_model}
The proposed D-IM system is designed for V2X communication environments consisting of two main components: 
\begin{itemize}
    \item \textbf{Blockchain Network:} A decentralized ledger that stores identity credentials (IDs, public keys, timestamps, location metadata, and cryptographic hashes). It serves as the immutable trust anchor for identity management. It consists of the \textbf{Verification Peers (RSUs)}. These verification validation peers hold the blockchain ledger records, maintain blockchain connectivity, assist in identity verification, and participate in consensus processes.
    \item \textbf{Participants Vehicles:} The primary participants that act as both data producers and consumers. Participants must be authenticated and verified before exchanging any information with each other or with the blockchain network.  We are planning to utilize machine learning based intrusion detection system \cite{abrar2024gps} and blockchain smart contracts to ensure fairness and penalize malicious behavior, revoke untrusted participants from the network and retract anomalous, false or misleading shared information as a future extension work based on this paper.
\end{itemize}

The proposed Zero Trust decentralized communication framework requires every identity in the network is verified before any data exchange. Vehicles and RSUs interact with the blockchain to obtain authenticated identity information. RSUs, acting as verification nodes, participate in validating credentials and anomaly detection. Suspicious behaviors are flagged, and verification nodes may collaborate to reach consensus on whether a participant is misbehaving.

\subsection{Attacker Model} \label{sec:attacker_model}
We assume an adversarial environment where attackers may attempt to compromise vehicular communications. Specific attack vectors include:
\begin{enumerate}
    \item \textbf{Identity and Authentication Attacks:}
    \begin{itemize}
        \item \textit{Masquerading/Impersonation Attack:} occurs when an attacker compromises a node (such as an Electronic Control Unit or an entire vehicle) and uses its legitimate identity to send malicious messages, making them appear genuine to other systems. The attacker then uses the gained access for eavesdropping or sending false sensors data, i.e. position/velocity data falsification.
        \item \textit{Sybil Attacks:} are when one attacker creates multiple fake identities (Sybil nodes) to broadcast false data (like fake accidents, traffic jams) to control or disrupt the network, compromising safety by creating unreliable traffic info or forcing bad routes. Attackers forge IDs by stealing legitimate ones or creating new ones, aiming to manipulate decisions in these decentralized systems, requiring strong authentication and behavior analysis for mitigation. 
    \end{itemize}
    \item \textbf{Data Integrity and Freshness Attacks:}
    \begin{itemize}
        \item \textit{Position/Velocity Data Falsification Attack:} where a malicious vehicle broadcasts false location/speed data within its basic safety message (BSM). This attack can mislead other vehicles, potentially leading to accidents or congestion by creating a false sense of their position. 
        \item \textit{Replay Attack:} involves an attacker capturing legitimate vehicle data and retransmitting it later to deceive the system, potentially causing accidents.
        \item \textit{Spoofing Attack:} involves a malicious actor sending fake signals to trick the vehicle's sensors, leading it to a false location or making it misinterpret its surroundings. Sometimes, this attack is preceded by an impersonation attack. Examples of such attack could be position/velocity data falsification attack, GPS spoofing, LiDAR spoofing or other sensors' data spoofing.
        \item \textit{Man-in-the Middle (MiTM) Attack:}  occurs when a malicious entity secretly intercepts, relays, and potentially alters communication between legitimate nodes, such as vehicles, or roadside units (RSUs). It can be passive, eavesdropping or active, in which  the attacker actively intervenes in the communication by dropping, delaying, or tampering with the content of messages, i.e. spoofing attack.
    \end{itemize}
\begin{figure}
\begin{center}
\includegraphics[clip,width=0.5\columnwidth]{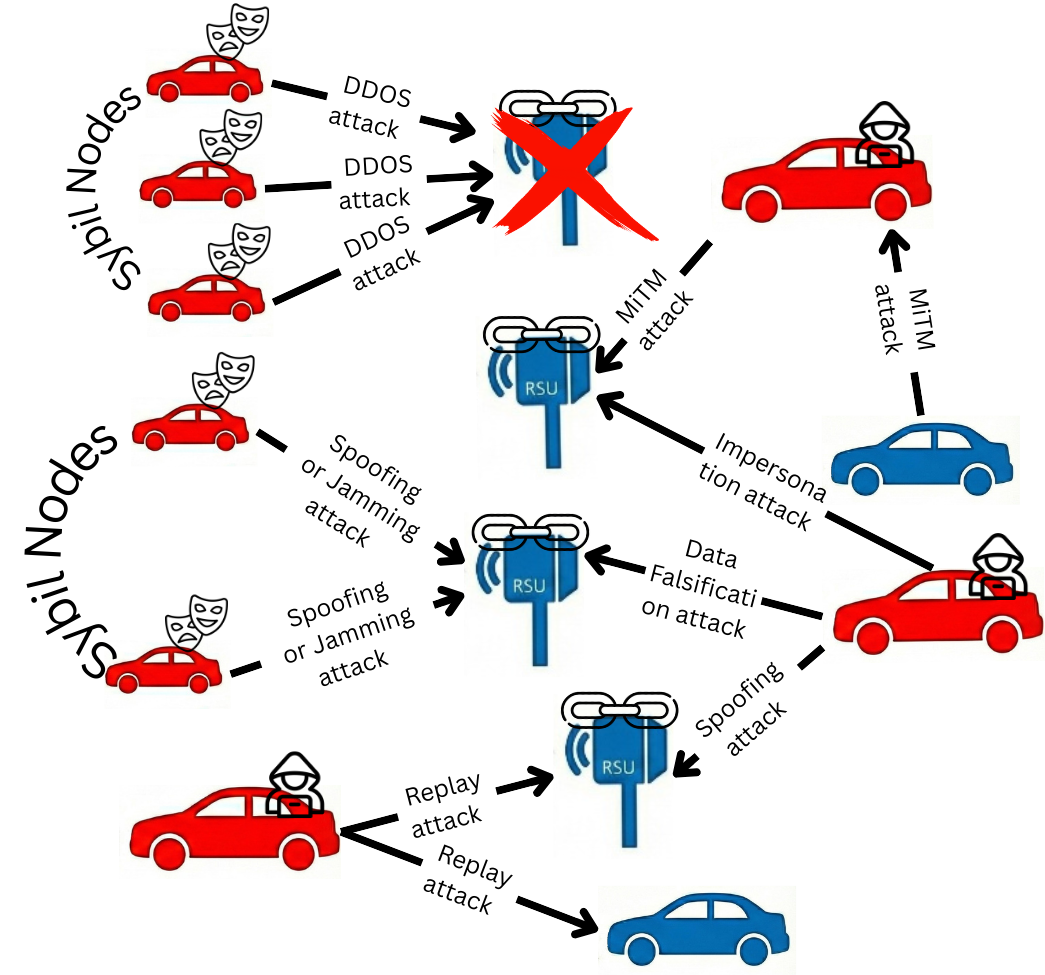}
\end{center}
\caption{Autonomous Vehicles Zero-Trust Decentralized Blockchain-based System Example Target Attacks}
\label{fig:attack_model}
\end{figure}
    \item \textbf{Collusion and Coordinated Attacks:}
    \begin{itemize}
        \item \textit{Deception and Misinformation:} Malicious vehicles or roadside units can collude to broadcast false information, such as forged warnings about obstacles or accidents, leading a victim vehicle to react incorrectly.
        \item \textit{Sensor Spoofing and Jamming:} Multiple vehicles can coordinate to spoof sensors, such as radar or GPS, simultaneously to confuse a target vehicle's perception system or jam its signals, creating a more robust and difficult-to-detect attack. 
        \item \textit{Adversarial Machine Learning:} Attackers can craft specific, coordinated inputs to confuse the machine learning models used for decision-making, such as altering perception data from multiple sensors to cause a misclassification.
    \end{itemize}

    \item \textbf{Availability and Network-Level Attacks:}
    \begin{itemize}
        \item \textit{Distributed Denial-of-Service (DDOS) Attack:} is a malicious attempt to disrupt the normal functionality and availability of blockchain network, vehicle, or RSU by overwhelming it with a flood of traffic from multiple coordinated sources.
    \end{itemize}

    \item \textbf{Accountability and Trust Attacks:}
    \begin{itemize}
        \item \textit{Repudiation Attacks:} happen when a system lacks strong controls to prove who initiated an action, allowing a malicious actor to deny sending a message or performing a transaction, or to alter logs to frame someone else, making digital records unreliable. The goal of such attacks could be denial of origin, log temparing or data forgery. 
        \item \textit{Single Point of Failure Attacks:} target a single component, like a critical sensor, communication link, or central server, that, if compromised or disabled, brings down the entire system or causes catastrophic failure, such as a crash or loss of control. Attackers exploit these vulnerabilities through sensor spoofing, V2X (Vehicle-to-Everything) exploits, or denial-of-service (DOS) attacks.
    \end{itemize}
\end{enumerate}

These attacks have been documented in prior work on vehicular networks and cooperative driving including \cite{yousseef2025autonomous,miller2015remote,amoozadeh2015security, satam2022autonomous}, motivating zero-trust based decentralized and continuously verifiable identity management solutions.  Figure \ref{fig:attack_model} illustrates some of these target attacks. The proposed D-IM framework mitigates these attacks through decentralized verification, hash-based integrity checks, and continuous authentication.

\section{DECENTRALIZED IDENTITY MANAGEMENT (D-IM) SYSTEM ARCHITECTURE} \label{sec:dim_design}

\begin{figure*}
\begin{center}
\includegraphics[clip,width=0.95\columnwidth]{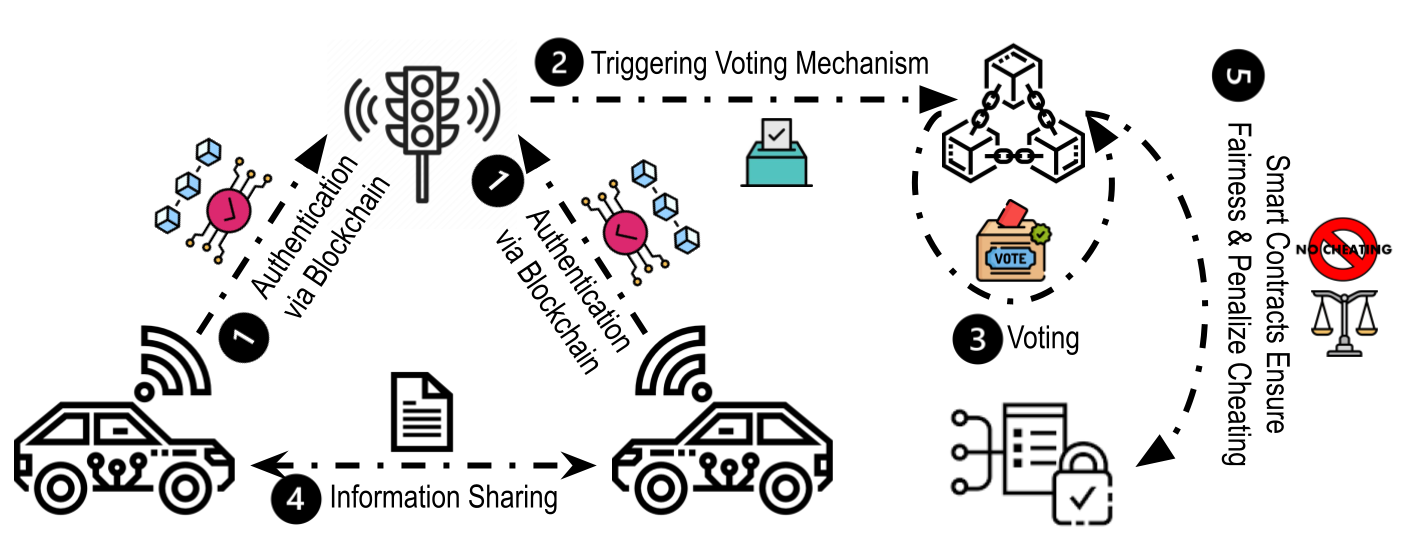}
\end{center}
\caption{System Steps of the proposed Zero-Trust Blockchain-based Communication Framework.}
\label{fig:sys_steps}
\end{figure*}
An overview of the system steps is shown in Figure~\ref{fig:sys_steps}. The overall communication framework integrates Zero Trust principles into V2X security. Authentication via blockchain precedes all information sharing, ensuring that only legitimate participants contribute to cooperative V2X operations. Each vehicle and RSU runs a local copy of the framework code and communicates with the blockchain to achieve secure and verifiable V2X communications. The framework enforces Zero Trust by requiring identity verification for every interaction, eliminating reliance on any single trusted entity. This paper focus on the authentication and authorization of joining vehicles (Step 1). The trust consensus voting and smart contract-based penalization (steps 3 and 5) are reserved for future work.

\begin{figure*}
\begin{center}
\includegraphics[clip,width=0.95\columnwidth]{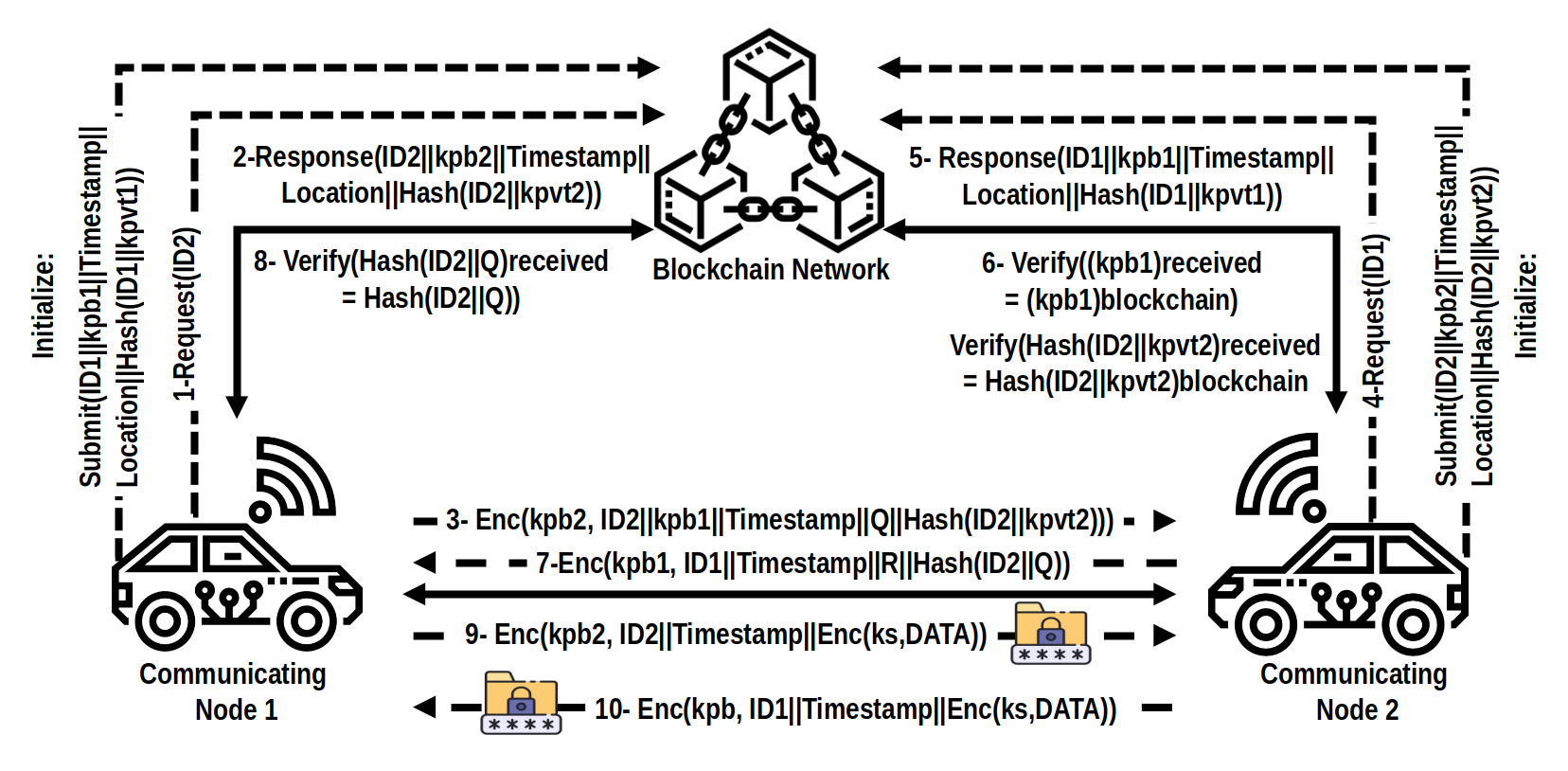}
\end{center}
\caption{Zero-Trust Decentralized Blockchain-based Identification and Authorization Protocol.}
\label{fig:ID_protocol}
\end{figure*}
The core of the D-IM system is a blockchain-based identification and authorization protocol, depicted in Figure~\ref{fig:ID_protocol}. The protocol allows vehicles to mutually authenticate, verify identity attributes through blockchain records, and establish secure communication channels without relying on centralized authorities.

\begin{table}[h]
  \caption{Algorithm \ref{alg:algorithm} used Parameters }
  \begin{tabular}{|c|c|}
    \hline
    \rowcolor[HTML]{EFEFEF} \textbf{Parameter} & \textbf{Explanation}  \\ \hline
    $ID_1$ & Vehicle~1 Pseudonymous Identity\\ \hline
    $ID_2$ & Vehicle~2 Pseudonymous Identity\\ \hline
    $TS$ & Timestamp\\ \hline
    $LOC$ & Vehicle's location\\ \hline
    $K_{pb_1}$ & Vehicle~1 Public Key\\ \hline
    $K_{pvt_1}$ & Vehicle~1 Private Key\\ \hline
    $K_{pb_2}$ & Vehicle~2 Public Key\\ \hline
    $K_{pvt_2}$ & Vehicle~2 Private Key\\ \hline
    $Q$ , $R$ & Random Numbers used to generate Diffie-Hellman based secret key\\ \hline
    $K_s$ & Diffie-Hellman Secret Key\\ \hline
  \end{tabular}%
  \label{tab:alg_parameters}
\end{table}

\begin{algorithm}[h]
\normalsize 
\caption{Zero Trust Blockchain-based Identification and Authorization Protocol}
\label{alg:algorithm}
\begin{algorithmic}[1]
\State \textbf{Vehicle$_1$} submits $(ID_1, K_{pb_1}, TS, Loc, Hash(ID_1||K_{pvt_1}))$ to \textbf{blockchain} \label{alg:step_1}
\State \textbf{Vehicle$_2$} submits $(ID_2, K_{pb_2}, TS, Loc, Hash(ID_2||K_{pvt_2}))$ to \textbf{blockchain} \label{alg:step_2}
\State \textbf{Vehicle$_1$} requests $ID_2$ info from blockchain \label{alg:step_3}
\State \textbf{blockchain} responds with $(ID_2, K_{pb_2}, TS, Loc, Hash(ID_2||K_{pvt_2}))$  \label{alg:step_4}
\State \textbf{Vehicle$_1$} sends to \textbf{Vehicle$_2$}: Enc$_{K_{pb_2}}(ID_1, ID_2, K_{pb_1}, TS, Q, Hash(ID_2||K_{pvt_2}))$ \label{alg:step_5}
\State \textbf{Vehicle$_2$} requests $ID_1$ info from \textbf{blockchain} \label{alg:step_6}
\State \textbf{blockchain} responds with $(ID_1, K_{pb_1}, TS, Loc, Hash(ID_1||K_{pvt_1}))$ \label{alg:step_7}
\If{($K_{pb1}$ matches blockchain record) AND (Hash($ID_2$||$K_{pvt2}$) matches blockchain record)} \label{alg:step_8}
    \State \textbf{Vehicle$_2$} sends to \textbf{Vehicle$_1$}: Enc$_{K_{pb_1}}(ID_2, ID_1, TS, R, Hash(ID_2||Q))$ \label{alg:step_9}
    \If{Hash($ID_2$||Q) received = locally computed Hash($ID_2$||Q)} \label{alg:step_10} 
        \State Both vehicles compute $K_s$ from $Q$ and $R$ via Diffie-Hellman \label{alg:step_11}
        \State Secure data exchange begins using $K_s$ \label{alg:step_12}
  \EndIf
\EndIf
\end{algorithmic}
\end{algorithm}

The detailed algorithm is provided in Algorithm~\ref{alg:algorithm} with its parameters summarized in Table \ref{tab:alg_parameters}. The algorithm formalizes the identification and authorization steps between communicating nodes. The protocol operates as follows:
\begin{enumerate}
    \item \textbf{Initialization:} Each vehicle submits its pseudonymous ID, public key, timestamp (to ensure message freshness), location (to verify physical proximity to nearby RSUs to ensure that a vehicle is physically present at its claimed location using received signal strength (RSS) measurements and to enable reliable association of reported events with a concrete geographic context), and the hash of its ID with its private key to the blockchain. The hash of the vehicle's ID and private key is used to verify the real identity of the vehicle since the public/private key pair is associated with real vehicle plate number or other identifying information. 
    \begin{itemize}
        \item \textit{Step 1:} \textbf{Vehicle$_1$} submits $(ID_1, K_{pb_1}, TS, Loc, Hash(ID_1||K_{pvt_1}))$ to \textbf{blockchain}.
        \item \textit{Step 2:} \textbf{Vehicle$_2$} submits $(ID_2, K_{pb_2}, TS, Loc, Hash(ID_2||K_{pvt_2}))$ to \textbf{blockchain}.
    \end{itemize}
    
    \item \textbf{Request/Response:} When Vehicle~1 requests to communicate with Vehicle~2, it queries the blockchain for Vehicle~2’s credentials and gets the blockchain response with Vehicle~2's credentials.
    \begin{itemize}
        \item \textit{Step 3:} \textbf{Vehicle$_1$} requests $ID_2$ info from blockchain.
        \item \textit{Step 4:} \textbf{blockchain} responds with $(ID_2, K_{pb_2}, TS, Loc, Hash(ID_2||K_{pvt_2}))$ 
    \end{itemize}
    
    Then, Vehicle~1 initiates communication with Vehicle~2 sending its own ID $ID_1$ to announce who is initiating communication, Vehicle~2's ID $ID_2$ for Vehicle~2 to know it is the intended recipient, its own public key $K_{pb_1}$ to be used for future communication, current timestamp $TS$, random number $Q$ to be used to generate Diffie-Hellman's secret key, and the hash $Hash(ID_2||K_{pvt_2}$ to verify Vehicle~1 is who claims it is since only the blockchain knows this hash and hence Vehicle~1 one must have be authorized by the blockchain to get the hash. The whole message is encrypted using Vehicle~2's public key $K_{pb_2}$.
     \begin{itemize}
        \item \textit{Step 5:} \textbf{Vehicle$_1$} sends to \textbf{Vehicle$_2$}: Enc$_{K_{pb_2}}(ID_1, ID_2, K_{pb_1}, TS, Q, Hash(ID_2||K_{pvt_2}))$
    \end{itemize}
    Similarly, Vehicle~2 queries the blockchain for Vehicle~1’s information and gets the blockchain response with Vehicle~2's credentials.
    \begin{itemize}
        \item \textit{Step 6:} \textbf{Vehicle$_2$} requests $ID_1$ info from \textbf{blockchain}
        \item \textit{Step 7:} \textbf{blockchain} responds with $(ID_1, K_{pb_1}, TS, Loc, Hash(ID_1||K_{pvt_1}))$
    \end{itemize}
    
    \item \textbf{Verification:} Vehicle~2 cross-verifies the Vehicle~1’s public key against blockchain records and received hash against its local records to ensure authenticity.
    \begin{itemize}
        \item \textit{Step 8:} {($K_{pb1}$ matches blockchain record) AND (Hash($ID_2$||$K_{pvt2}$) matches blockchain record)} 
    \end{itemize}
    If the received key and hash match, Vehicle~2 sends to Vehicle~1 its own ID $ID_2$ to announce who is initiating communication, Vehicle~1's ID $ID_1$ for Vehicle~1 to know it is the intended recipient, its own public key $K_{pb_2}$, current timestamp $TS$, random number $R$ to be used to generate Diffie-Hellman's secret key, and the hash $Hash(ID_2||Q))$ to be used to verify Vehicle~2 received the right $Q$ value. The whole message is encrypted using Vehicle~1's public key $K_{pb_1}$.
    \begin{itemize}
        \item \textit{Step 9:} \textbf{Vehicle$_2$} sends to \textbf{Vehicle$_1$}: Enc$_{K_{pb_1}}(ID_2, ID_1, TS, R, Hash(ID_2||Q))$ 
    \end{itemize}
    Vehicle~1 cross-verifies received hash against its local computed hash to ensure authenticity.
    \begin{itemize}
        \item \textit{Step 10:} Hash($ID_2$||Q) received = locally computed Hash($ID_2$||Q)
     \end{itemize}
    \item \textbf{Session Key Establishment:} Both vehicles have exchanged random values ($Q$, $R$) encrypted with the other’s public key, enabling them to compute a shared session key $K_s$ using the Diffie-Hellman method. 
    \begin{itemize}
        \item \textit{Step 11:} Both vehicles compute $K_s$ from $Q$ and $R$ via Diffie-Hellman
     \end{itemize}
    \item \textbf{Secure Communication:} Subsequent information exchange is encrypted with the established session key $K_s$, ensuring confidentiality and integrity.  
    \begin{itemize}
        \item \textit{Step 12:} Secure data exchange begins using $K_s$
     \end{itemize}
      
\end{enumerate}

\section{SECURITY ANALYSIS} \label{sec:sec_analysis}
\subsection{Formal BAN Logic-Based Security Analysis}
To formally demonstrate the security properties of the proposed Zero Trust-based Decentralized Identity Management, D-IM, system, we employ Burrows-Abadi-Needham (BAN) logic \cite{UCAM-CL-TR-138, burrows1990logic} to provide a formal proof  that our system goals have been achieved. BAN logic is a formal security hypothesis based security analysis used to prove the validity of securiry assumptions. BAN logic has been extensively applied to validate mutual authentication, key agreement, and freshness guarantees in security-critical systems, including vehicular and wireless architectures \cite{abdel2022proxy, farrea2025zero, guo2024uava}. BAN logic allows us to model how two vehicles evolve their beliefs about each other’s identity and about the established session key through the exchange of protocol messages, under clearly specified assumptions.

Table~\ref{tab:ban_notations} summarizes the BAN logic notations used throughout this analysis.

\begin{table}[h]
\centering
\caption{BAN Logic Notations}
\label{tab:ban_notations}
\begin{tabular}{ll}
\hline
\textbf{Notation} & \textbf{Description} \\
\hline
$X \mid\equiv M$ & Entity $X$ believes that statement $M$ is true \\
$X \triangleleft M$ & Entity $X$ has received message $M$ \\
$X \mid\sim M$ & Entity $X$ once said message $M$ \\
$X \Rightarrow M$ & Entity $X$ has jurisdiction over statement $M$ \\
$\#(M)$ & Message $M$ is fresh \\
$X \leftrightarrow_K Y$ & $X$ and $Y$ share a secret key $K$ \\
$K_{pbX}$ & Public key of entity $X$ \\
$K_{pvX}$ & Private key of entity $X$ \\
\hline
\end{tabular}
\end{table}

The BAN logic inference rules applied in this work are listed in
Table~\ref{tab:ban_rules}.

\begin{table}[h]
\centering
\caption{BAN Logic Inference Rules}
\label{tab:ban_rules}
\begin{tabular}{ll}
\hline
\textbf{Rule} & \textbf{Description} \\
\hline
\textit{Message Meaning} & 
$
\frac{
X \mid\equiv K_{pbY} ,
\quad X \triangleleft \{M\}_{K_{pvY}}
}{
X \mid\equiv Y \mid\sim M
}
$ \\[10pt]

\textit{Nonce Verification} &
$
\frac{
X \mid\equiv \#(M) ,
\quad X \mid\equiv Y \mid\sim M
}{
X \mid\equiv Y \mid\equiv M
}
$ \\[10pt]

\textit{Jurisdiction} &
$
\frac{
X \mid\equiv Y \Rightarrow M ,
\quad X \mid\equiv Y \mid\equiv M
}{
X \mid\equiv M
}
$ \\

\hline
\end{tabular}
\end{table}

In the D-IM setting, the main participants are Vehicle~1 ($V_1$) and Vehicle~2 ($V_2$), which mutually authenticate and establish a shared session key $K_s$ using the blockchain as a decentralized trust anchor. Blockchain verification peers (RSUs) maintain the ledger, but from the BAN perspective they are abstracted as a single trusted authority that certifies public keys and identity bindings. The analysis focuses on the identification and authorization protocol (Algorithm \ref{alg:algorithm}), where $V_1$ and $V_2$ verify each other’s credentials via the blockchain and then perform a Diffie-Hellman-based key exchange. We denote the blockchain authority (the set of verification peers / RSUs that maintain the ledger) as $BC$ for the purpose of initial trust assumptions.

\subsubsection{Security Goals and Assumptions}
In accordance with Zero Trust principles, the D-IM protocol does not rely on implicit trust and instead enforces continuous verification and explicit belief derivation. The following security goals are defined for the interaction between $V_1$ and $V_2$:

\begin{itemize}
    \item \textbf{SG1:} $V_1 \mid\equiv V_2$ \\
    Vehicle $V_1$ believes that it is communicating with a legitimate $V_2$.
    \item \textbf{SG2:} $V_2 \mid\equiv V_1$ \\
    Vehicle $V_2$ believes that it is communicating with a legitimate $V_1$.
    \item \textbf{SG3:} $V_1 \mid\equiv V_1 \leftrightarrow_{K_s} V_2$ \\
    Vehicle $V_1$ believes that it shares a fresh session key $K_s$ with $V_2$.
    \item \textbf{SG4:} $V_2 \mid\equiv V_1 \leftrightarrow_{K_s} V_2$ \\
    Vehicle $V_2$ believes that it shares a fresh session key $K_s$ with $V_1$.
\end{itemize}

The following initial assumptions are derived from the design of the D-IM system:

\begin{itemize}
    \item \textbf{IA1:} $V_1 \mid\equiv BC \Rightarrow (ID_2, K_{pb2}, Hash(ID_2||K_{pvt_2}))$ \\
    $V_1$ believes that $BC$ has jurisdiction over $V_2$’s identity, public key, and corresponding hash.
    \item \textbf{IA2:} $V_2 \mid\equiv BC \Rightarrow (ID_1, K_{pb1}, Hash(ID_1||K_{pvt_1})))$\\
    $V_2$ believes that $BC$ has jurisdiction over $V_1$’s identity, public key, and corresponding hash.
    \item \textbf{IA3:} $V_1 \mid\equiv \#(Q)$\\
    $V_1$ believes that its Diffie-Hellman contribution $Q$ is fresh.
    \item \textbf{IA4:} $V_2 \mid\equiv \#(R)$\\
    $V_2$ believes that its Diffie-Hellman contribution $R$ is fresh.
    \item \textbf{IA5:} $V_1 \mid\equiv K_{pb2}$ is the authentic public key of $V_2$ since it got it from $BC$.
    \item \textbf{IA6:} $V_2 \mid\equiv K_{pb1}$ is the authentic public key of $V_1$ since it got it from $BC$.
    \item \textbf{IA7:} $V_1 \mid\equiv V_2 \Rightarrow (V_1 \leftrightarrow_{K_s} V_2)$\\
    $V_1$ believes $V_2$ has jurisdiction over statements about the shared session key with $V_1$.
    \item \textbf{IA8:} $V_2 \mid\equiv V_1 \Rightarrow (V_1 \leftrightarrow_{K_s} V_2)$\\
    $V_2$ believes $V_1$ has jurisdiction over statements about the shared session key with $V_2$.
\end{itemize}
These assumptions reflect that (i) public keys and identity bindings are certified by the blockchain, (ii) Diffie-Hellman contributions are fresh, and (iii) each vehicle trusts the other party to control statements about the shared session key.

\subsubsection{D-IM BAN Logic-based Security Analysis}
The core messages of our proposed identification and authorization protocol (Algorithm~1) are as follows:
\begin{itemize}
    \item \textbf{Credential retrieval via blockchain}
        \begin{align*}
            V_1 &\triangleleft (ID_2, K_{pb2}, Hash(ID_2||K_{pvt_2}))) \text{ from } BC, \\
            V_2 &\triangleleft (ID_1, K_{pb1}, Hash(ID_1||K_{pvt_1}))) \text{ from } BC.
        \end{align*}
    After this step, using IA1--IA2 and the jurisdiction rule, both vehicles believe that the retrieved identity, public key and hash information are correct.
    \item \textbf{Vehicle~1 $\rightarrow$ Vehicle~2}: $M_1 : V_2 \triangleleft \{ ID_2, K_{pb1}, TS, Q, Hash(ID_2||K_{pvt_2})) \}_{K_{pb2}}$
    $V_1$ sends its Diffie--Hellman value $Q$ and related information encrypted with $V_2$'s public key.
    \item \textbf{Vehicle~2 $\rightarrow$ Vehicle~1}: $M_2 : V_1 \triangleleft \{ ID_1, T_S, R, \mathsf{Hash}(ID_2 \parallel Q) \}_{K_{pb1}}$
    $V_2$ responds with its Diffie--Hellman value $R$ and a hash binding $ID_2$ and $Q$, encrypted with $V_1$'s public key.
    \item \textbf{Session key derivation}
        Both vehicles, $v_1$, $V_2$ compute: $ K_s = f(Q,R)$ using Diffie--Hellman. Subsequent messages are encrypted under $K_s$.
\end{itemize}

We now sketch how SG1--SG4 are derived using standard BAN rules.
\begin{itemize}
    \item \textbf{From $M_1$ at $V_2$.}
    \begin{itemize}
        \item From $M_1$, $V_2$ receives: $V_2 \triangleleft \{ ID_2, K_{pb1}, TS, Q, Hash(ID_2||K_{pvt_2}))\}_{K_{pb2}}$
        \item Using IA6 (authenticity of $K_{pb2}$) and the message meaning rule, $V_2$ infers: $ V_2 \mid\equiv V_1 \mid\sim (ID_2, K_{pb1}, T_S, Q, Hash(ID_2||K_{pvt_2}))).$
        \item Using IA3 ($\#(Q)$) and the nonce verification rule, $V_2$ derives: $ V_2 \mid\equiv V_1 \Rightarrow Q$
        \item From the belief rule, $V_2$ thus believes that $V_1$ is alive and participating in the current protocol instance. Combined with the integrity of the blockchain bindings (IA2) and consistency of $H_2$, $V_2$ concludes that the peer is the legitimate $V_1$, leading toward \textbf{SG2}:
            \begin{equation*}
                V_2 \mid\equiv V_1.
            \end{equation*}
    \end{itemize}
    
    \item \textbf{From $M_2$ at $V_1$.}
    \begin{itemize}
        \item From $M_2$, $V_1$ receives: $V_1 \triangleleft \{ ID_1, T_S, R, \mathsf{Hash}(ID_2 \parallel Q) \}_{K_{pb1}}$
       \item Using IA5 (authenticity of $K_{pb1}$) and the message meaning rule, $V_1$ infers: $V_1 \mid\equiv V_2 \mid\sim (ID_1, T_S, R, \mathsf{Hash}(ID_2 \parallel Q))$ 
       \item Using IA4 ($\#(R)$) and the nonce verification rule, $V_1$ derives: $V_1 \mid\equiv V_2 \Rightarrow R$
        \item Since the received hash $\mathsf{Hash}(ID_2 \parallel Q)$ matches $V_1$'s locally computed value, $V_1$ is assured that $V_2$ correctly received and used $Q$, binding the current run of the protocol to the same peer. Together with the blockchain-confirmed credentials (IA1), this implies \textbf{SG1}:
            \begin{equation*}
                V_1 \mid\equiv V_2.
            \end{equation*}
    \end{itemize}
    
    \item \textbf{Session key beliefs.}
    \begin{itemize}
        \item Both parties compute the same session key $K_s = f(Q,R)$. From their beliefs in the freshness of $Q$ and $R$ (IA3--IA4) and their belief that these values were correctly used by the peer, each party concludes that $K_s$ is fresh:
            \begin{align*}
                V_1 \mid\equiv \#(K_s), \\
                V_2 \mid\equiv \#(K_s).
            \end{align*}
        \item Using IA7 and IA8 (jurisdiction over shared-key statements) and the jurisdiction rule, we derive:
            \begin{align*}
                V_1 \mid\equiv V_1 \leftrightarrow_{K_s} V_2 &\quad\textbf{(SG3)}, \\
                V_2 \mid\equiv V_1 \leftrightarrow_{K_s} V_2 &\quad\textbf{(SG4)}.
            \end{align*}
    \end{itemize}
\end{itemize}
Thus, the BAN logic derivation shows that both vehicles mutually authenticate each other and agree on a fresh session key $K_s$ that is known only to them, under the stated assumptions.\\

The BAN logic analysis confirms that the D-IM identification and authorization protocol achieves its core Zero Trust-aligned security goals: mutual authentication, fresh key establishment, and resistance to impersonation and replay, assuming the blockchain authority correctly certifies public keys and that the underlying cryptographic primitives are secure. In other words, if an attacker cannot subvert the blockchain’s credential records or break the underlying cryptography, then it cannot successfully impersonate a vehicle or derive the established session key within the modeled threat scope.

\subsection{Qualitative Security Analysis} \label{sec:informal_sec_anaylsis}
This subsection provides a qualitative security analysis of the proposed Zero Trust-based D-IM system, focusing on how the design mitigates the attacks defined in the attacker model subsection \ref{sec:attacker_model}. The goal is to clearly illustrate the security rationale behind the framework and its alignment with Zero Trust principles. In line with the “never trust, always verify” philosophy, the system enforces continuous authentication, decentralized verification, and tamper-evident identity binding for all V2X entities. As a result, several attacks are prevented by design, while others are detected, contained, or their impact is significantly reduced. We also distinguish between guarantees provided by the current implementation and additional protection achievable through planned extensions such as anomaly-based detection and voting-based consensus. 
\begin{enumerate}
   \item \textbf{Identity and Authentication Attacks}
    \begin{itemize}
        \item \textbf{Masquerading / Impersonation Attacks:}
        Impersonation attacks rely on an attacker assuming the identity of a legitimate vehicle or RSU to inject malicious messages. In the proposed D-IM system, each entity’s identity is cryptographically bound to a public key and registered on a permissioned blockchain. Mutual authentication requires cross-verification of blockchain-stored credentials before any communication is established. As a result, an attacker lacking the corresponding private key cannot successfully assume another entity’s identity. This attack is therefore prevented by design in the current implementation. Revocation records stored immutably on the blockchain further ensure that compromised identities cannot regain trust after detection.
        \item \textbf{Sybil Attacks:}
        Sybil attacks attempt to introduce multiple fake identities to influence cooperative decisions. The D-IM framework mitigates this threat through decentralized identity registration and verification, where each identity must be validated by blockchain verification peers (RSUs) before being accepted. The cost and traceability of identity creation significantly raise the barrier for generating large numbers of pseudonymous identities. This provides strong prevention against large-scale Sybil attacks. Future extensions, such as behavior-based voting and reputation management, can further limit residual Sybil influence by correlating identity behavior across time.
    \end{itemize}

    \item \textbf{Data Integrity and Freshness Attacks}
    \begin{itemize}
        \item \textbf{Position/Velocity Data Falsification:} 
        In this attack, a legitimate but malicious participant broadcasts incorrect kinematic data. While identity verification alone cannot guarantee data correctness, the D-IM system ensures that falsified data is always attributable to a verified sender. This provides accountability and detection, rather than full prevention. Hash-based integrity checks and authenticated session keys prevent in-transit message tampering, while the Zero Trust enforcement ensures that only authenticated entities can publish messages. Planned integration with anomaly-based behavior analysis can further identify implausible trajectories or inconsistent motion patterns and limit their influence. Also, since each vehicle shares its location,  during both authentication and periodic status updates, with the blockchain network residing on RSUs, RSUs can verify the reported location accuracy using the received signal strength (RSS) of corresponding received messages.

        \item \textbf{Replay Attacks:} 
        Replay attacks involve re-transmitting previously valid messages to mislead receivers. The proposed protocol incorporates timestamps and freshness checks during identity verification and message exchange. Messages with stale timestamps or reused cryptographic challenges are rejected during the authentication handshake. Consequently, replay attacks are prevented by design in the current implementation.

        \item \textbf{Sensor Spoofing (GPS, LiDAR, Radar):}
        Sensor spoofing targets the perception layer and cannot be fully prevented by identity management alone. However, the D-IM framework limits its impact by ensuring that external data used in cooperative perception originates from authenticated and authorized entities. This enables detection and containment, as inconsistent data from authenticated senders can be flagged and correlated across multiple sources. Full mitigation is strengthened through future extensions involving multi-source validation and anomaly-based intrusion detection.
        \item \textbf{Man-in-the-Middle, MiTM, Attacks:}
        MiTM attacks attempt to intercept or modify communication between legitimate entities. In the D-IM protocol, all session establishment occurs after mutual blockchain-backed identity verification and results in an encrypted session key derived via Diffie-Hellman exchange. Without access to the private keys, an intermediary cannot decrypt or alter messages without detection. As a result, active MiTM attacks are prevented, while passive eavesdropping is rendered ineffective by encryption.
   \end{itemize}

   \item \textbf{Collusion and Coordinated Attacks}
   \begin{itemize}
       \item \textbf{Deception and Coordinated Misinformation:}
       Colluding vehicles may attempt to jointly broadcast false information (e.g., fake hazards). The D-IM system limits the impact of such attacks by ensuring that all data is attributable and traceable to verified identities. While collusion cannot be entirely prevented at the identity layer, the framework enables detection, attribution, and post-event accountability. Planned consensus-based validation and voting among RSUs can further reduce the influence of coordinated false reports.
       \item \textbf{Adversarial Machine Learning Attacks:}
       These attacks target downstream ML models rather than the communication substrate itself. While D-IM does not directly prevent adversarial manipulation of perception models, it restricts ML inputs to authenticated data sources and preserves data integrity during transmission since only authenticated vehicles can participate in communication and data sharing. This reduces the attack surface and enables more effective detection when combined with future ML-based intrusion detection systems.
    \end{itemize}

    \item \textbf{Availability and Network-Level Attacks}
    \begin{itemize}
        \item \textbf{Distributed Denial-of-Service (DDoS) Attacks:}
        DDoS attacks aim to disrupt availability by overwhelming vehicles, RSUs, or the blockchain network. The permissioned nature of the blockchain and identity-gated access control restrict participation to authenticated entities, thereby reducing the attack surface. While high-volume DDoS attacks cannot be fully prevented, the framework provides attack containment, as malicious senders can be identified, rate-limited, and revoked. Future smart-contract-based penalties can further automate this response.
    \end{itemize}

    \item \textbf{Accountability and Trust Attacks}
    \begin{itemize}
        \item \textbf{Repudiation Attacks:}
        Repudiation occurs when a malicious entity denies having sent a message or performed an action. The immutable blockchain ledger records identity registration, key usage, and verification events, providing non-repudiable evidence. As a result, repudiation attacks are prevented, as entities cannot plausibly deny authenticated actions.
        \item \textbf{Single Point of Failure Exploitation:}
        Traditional centralized identity systems suffer from single points of failure. By decentralizing identity verification and credential storage across multiple RSUs and blockchain peers, the D-IM framework eliminates reliance on any single trusted authority. This attack class is therefore prevented by design, with redundancy and decentralization preserving system availability even under partial compromise.
    \end{itemize}
\end{enumerate}
Overall, the proposed D-IM system leverages blockchain and Zero Trust principles to enable decentralized, tamper-resistant, and continuously verifiable identity management in V2X environments. The framework provides strong preventive guarantees against identity-based, replay, impersonation, man-in-the-middle, repudiation, and single-point-of-failure attacks, while enabling effective detection, attribution, and containment of data falsification, spoofing, collusion, and availability-based attacks. By ensuring rigorous authentication, efficient key establishment, and scalable operation, the proposed D-IM system establishes a robust security foundation for autonomous vehicle communication networks and successfully addresses the defined threat model. Future work will further strengthen this foundation through the integration of smart contracts and anomaly-based machine learning techniques to automate misbehavior penalties and enhance system adaptability.

\section{EXPERIMENTAL EVALUATION} \label{sec:sysEval}
This section evaluates the performance and practicality of the proposed D-IM system through simulation-based experiments in realistic V2X scenarios. The evaluation focuses on quantifying the communication overhead introduced by continuous identity verification, as well as its impact on key network performance metrics under varying traffic densities. The results demonstrate that the proposed framework provides strong security and attribution guarantees while maintaining scalability and efficiency suitable for real-world autonomous vehicle deployments.

\subsection{Simulation Setup }
\begin{table}[h]
\caption{Main simulation parameters and settings}
\centering
\large
\begin{tabular}{|ll|}
\hline
\textbf{Settings}             &                     \\ \hline
Beacon periodicity            & 10 Hz               \\
Beacon size                   & 1670 or 90 B        \\ \hline
\textbf{PHY Layer}            &                     \\ \hline
Channel bandwidth             & 10 MHz              \\
Transmission power            & 23 dBm              \\
Antenna gain (tx/rx)          & 3 dB                \\
Noise figure                  & 9 dB                \\
Propagation model             & WINNER+             \\
Shadowing variance            & LOS 3 dB, NLOS 4 dB \\ \hline
\textbf{DSRC Parameters}      &                     \\ \hline
Carrier sensing sensitivity   & -85 dBm             \\
Contention window             & 15                  \\ \hline
\textbf{C-V2X Parameters}     &                     \\ \hline
$p_k$ (resource retention)    & 0.8            \\
Sensing threshold             & -126 dBm            \\ \hline
\end{tabular}%
\label{table:settings}
\end{table}
This subsection presents the implementation details of our proposed identification and authorization protocol, and the simulation environment. The blockchain infrastructure is built on \textbf{Hyperledger Iroha v2} \cite{Iroha}, a permissioned blockchain designed for identity-centric applications, offering a lightweight and modular architecture well suited for distributed identity management in vehicular networks. Iroha employs the Yet Another Consensus (YAC) protocol, a crash-fault-tolerant consensus mechanism that prioritizes low latency and high throughput over full Byzantine resilience. , YAC achieves eventual consistency across network nodes while maintaining system resilience in the event of node failures \cite{Iroha_docs}. This design choice enables efficient transaction confirmation and makes Iroha a practical solution for consortium-based vehicular environments where participants are semi-trusted and real-time performance is critical.



The proposed identification and authorization protocol was implemented in Python. Communication with the Iroha blockchain was facilitated through the official Iroha Python library v0.0.5.5. Cryptographic operations were realized using the cryptography package v3.4.8, which provides RSA and symmetric key primitives, while Python’s secrets module generated random secret values for Diffie-Hellman key exchange. For vehicle-to-vehicle (V2V) communication, we employed the Python socket module to establish and manage encrypted communication channels between participants.

To evaluate the proposed protocol, we used \textbf{WiLabV2XSim v6.2} a discrete-event simulation framework developed in MATLAB for modeling vehicular networks with a focus on cooperative awareness services. WiLabV2XSim supports both DSRC and C-V2X communications and can incorporate synthetic mobility models or real traffic traces to represent vehicle movements \cite{bazzi2019survey}. We ran our simulations on High Performance Computing (HPC) resources supported by the University of Arizona. We dedicated one node with one core for each scenario (Different simulation scenarios explained in \ref{sec:scenarios}). Standard partition on Puma cluster was used which uses  CPU only. We used 6 GBytes memory per CPU. Matlab r2023b was used to run the WiLabV2XSim simulator.

The main simulation parameters are summarized in Table~\ref{table:settings}. For DSRC, channel access is modeled using carrier-sense multiple access with collision avoidance (CSMA/CA), while C-V2X employs sensing-based semi-persistent scheduling (SPS). These settings allow evaluation of the trade-offs between communication overhead introduced by the proposed protocol and the reliability, timeliness, and channel efficiency of vehicular communication in both highway and urban environments.

\subsection{Evaluation Parameters and Scenarios}\label{sec:scenarios}

\begin{figure}
\begin{center}
\includegraphics[clip,width=0.3\columnwidth]{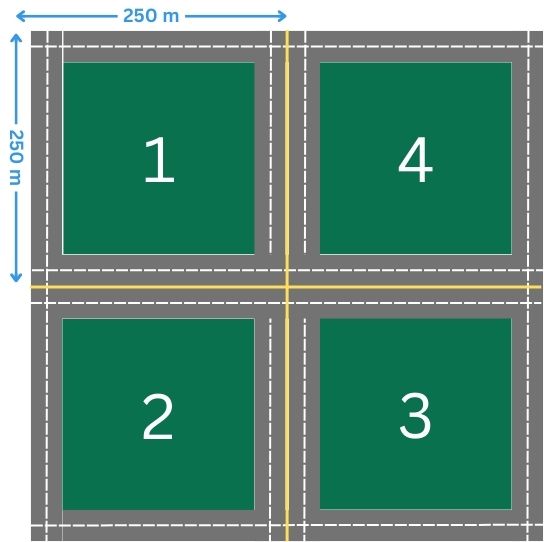}
\end{center}
\caption{Urban Environment Layout used}
\label{fig:urban_layout}
\end{figure}

Each simulation lasted one minute and was run for both highway and urban scenarios. In the highway case, vehicles traveled along a 5000~m bidirectional road with four lanes per direction, an average speed of 120~km/h, and a standard deviation of 20~km/h. In the urban case, traffic was modeled on a $2 \times 2$ grid comprising four blocks, each 250~m long and wide, with four lanes in both vertical and horizontal directions. Figure \ref{fig:urban_layout} shows the urban scenario layout used in our simulations. For both the highway and the urban scenarios, the lane width is 4 meters.The mean speed was 60~km/h with a standard deviation of 15~km/h.  Simulations were run using DSRC and LTE-V2X communication protocols sending either the minimum sized packet or the average sized packet. We ran the simulations with different vehicle densities in both the highway and the urban scenarios. Simulations for each combination of environment, communication protocol, packet size and vehicle density were repeated 10 times with different starting seed and the outputs were averaged. The framework was evaluated using four output parameters: 
\begin{itemize}
    \item \textbf{Packet Reception Rate (PRR):} Ratio of successfully decoded beacons by significant neighbors to the total transmitted beacons.  
    \item \textbf{Update Delay or Inter-Packet Gap (IPG):} Time between two consecutive successfully received beacons from the same node within the awareness range.  
    \item \textbf{Packet Age:} Time between the generation of a packet and its effective transmission.  
    \item \textbf{Channel Busy Ratio (CBR):} Channel occupancy quantification: for DSRC, it is calculated every $T_{CBR}=100$~ms as the ratio $T_{Busy}/T_{CBR}$, where $T_{Busy}$ is the time the channel is sensed busy; for C-V2X, it is computed as $N_{Busy}/N_{CBR}$, where $N_{Busy}$ is the number of subchannels marked busy and $N_{CBR}$ is the total number of subchannels within $T_{CBR}$.
\end{itemize}

\begin{figure*}[t]
\begin{center}
\subfloat[]{  
\includegraphics[width=0.5\columnwidth]{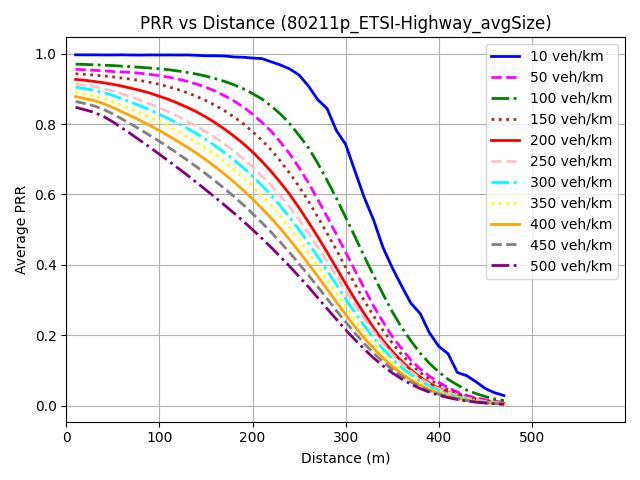}
  \label{fig:PRR_highway_dsrc_avg}}
\subfloat[]{
  \includegraphics[width=0.5\columnwidth]{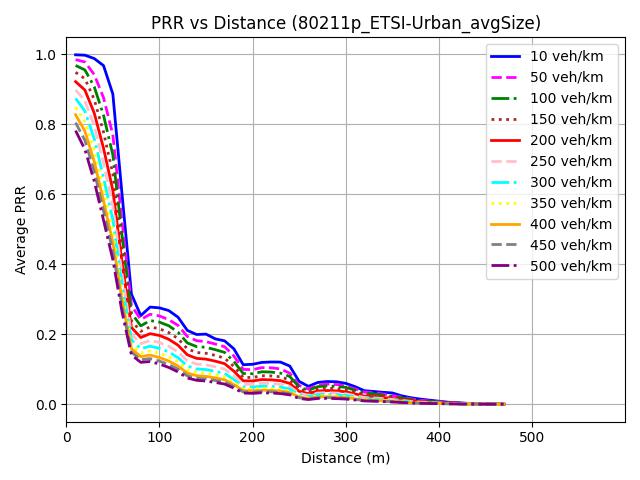}
  \label{fig:PRR_urban_dsrc_avg}}
  \vfill
\subfloat[]{  
\includegraphics[width=0.5\columnwidth]{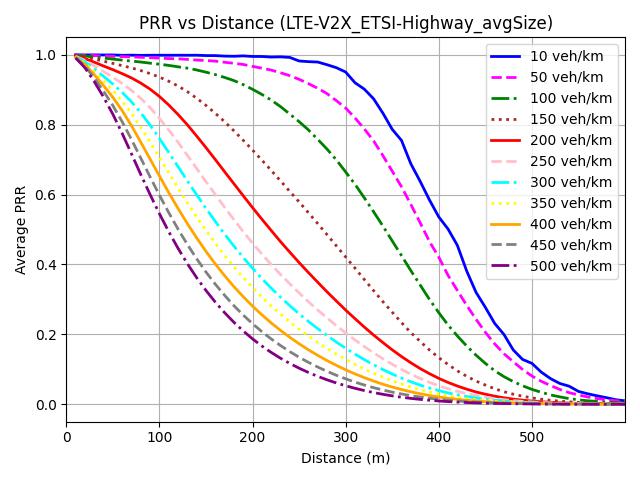}
  \label{fig:PRR_highway_lte_avg}}
\subfloat[]{
  \includegraphics[width=0.5\columnwidth]{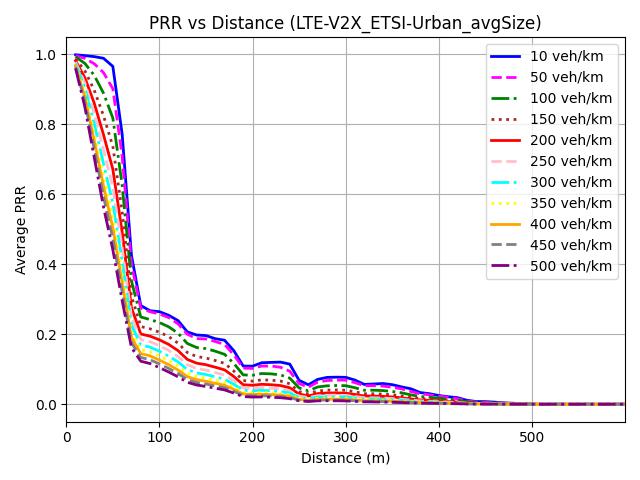}
  \label{fig:PRR_urban_lte_avg}}
\end{center} 
\caption{ (a) Highway Packet Reception Rate and (b) Urban Packet Reception Rate vs. the distance using DSRC RAT and average packet size required by our implementation (c) Highway Packet Reception Rate and (d) Urban Packet Reception Rate vs. the distance using LTE-V2X RAT and average packet size required by our implementation.} 
\label{fig:PRR_vs_dist}
\end{figure*}

\subsection{Simulation Results}
This section presents and analyzes the results of our simulation-based performance evaluation and presents a scalability analysis increasing the vehicle density for each scenario. The objective is to quantify the communication overhead introduced by our Zero Trust D-IM protocol, and assess its impact on critical vehicular network metrics such as PRR, CBR, and latency, and its scalability potential. The following figures present the key findings across various scenarios, demonstrating that our framework provides strong security guarantees with minimal performance trade-offs suitable for real-world deployment.
\begin{figure*}[ht]
\begin{center}
\subfloat[]{
  \includegraphics[width=0.5\columnwidth]{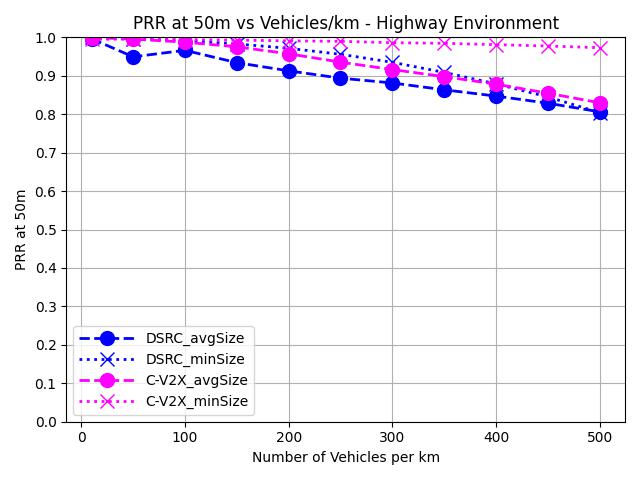}
  \label{fig:PRR_vehicles_highway}}
  \subfloat[]{
  \includegraphics[width=0.5\columnwidth]{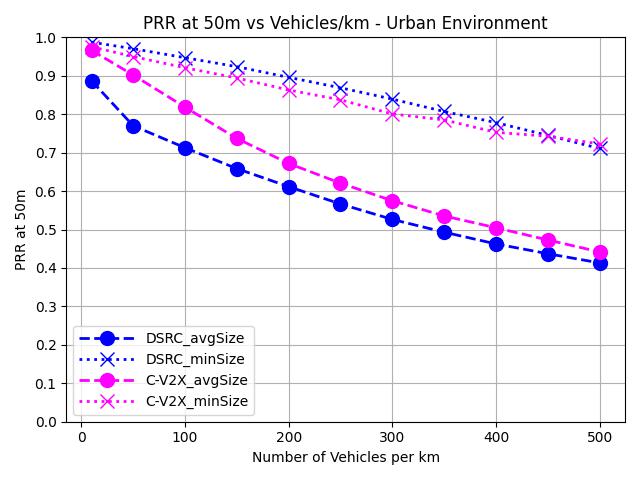}
  \label{fig:PRR_vehicles_urban}}
\end{center} 
\caption{ (a) Highway Packet Reception Rate and (b) Urban Packet Reception Rate vs. the Number of Vehicles per km at 50m.} 
\label{fig:PRR_vehicles}
\end{figure*}

Figure \ref{fig:PRR_vs_dist} shows the PRR vs distance with different communication protocols, DSRC AND C-V2X, for both urban and highway environments. The curves show that increasing the number of vehicles per km results in decreasing the PRR. Also, for a given number of vehicles per km, the PRR decreases with increasing the distance. Moreover, from the figure, we clearly find that PRR is higher in highway scenario than it is for urban scenario. This can be explained by the fact that in urban scenarios there are many obstacles, i.e. buildings that hinder the reception of packets. We, also, observe that the PRR is lower when DSRC is used than when LTE-V2X is used. This is because DSRC uses carrier sense multiple access/ collision avoidance (CSMA/CA) for media access which implements "listen-before-talk" strategy, while LTE-V2X mode 4, that enables peer-to-peer communication adopted in our simulations, uses Sensing-Based Semi-Persistent Scheduling (SB-SPS) for media access. SB-SPS also uses a "listen-before-talk" strategy and a periodic reservation system to manage channel access and reduce collisions. In SB-SPS, once a resource is selected, the vehicle reserves it for a specified number of consecutive transmissions at a regular interval. This semi-persistent reservation is more efficient than a full random access protocol for every packet resulting in higher PRR.

Our implementation of the zero trust blockchain based identification protocol poses an overhead in terms of the size of packets exchanged. To compare PRR when the average size required by our implementation is used for the simulation packet size to the case in which  the simulation packet size is set to the minimum packet size, we refer to Figure \ref{fig:PRR_vehicles} which presents the PRR vs the vehicle density per km with different communication protocols for both urban and highway environments. The graphs show that this overhead results in an average reduction of around 18\% in PRR for urban scenarios and reduction of 6\% in PRR for highway scenario using LTE-V2X RAT. For vehicle density of 200 vehicles/km, the reduction observed was of around 19\% in PRR for urban scenarios and  around 3\% in PRR for highway scenario using LTE-V2X RAT. For the cases where DSRC RAT is used, a average reduction of around 27\% in PRR for urban scenarios and average reduction of 4\% in PRR for highway scenarios. For vehicle density of 200 vehicles/km, the reduction observed was of around 28\% in PRR for urban scenarios and  around 6\% in PRR for highway scenario using DSRC RAT.
\begin{figure*}[ht]
\begin{center}
\subfloat[]{  
\includegraphics[width=0.5\columnwidth]{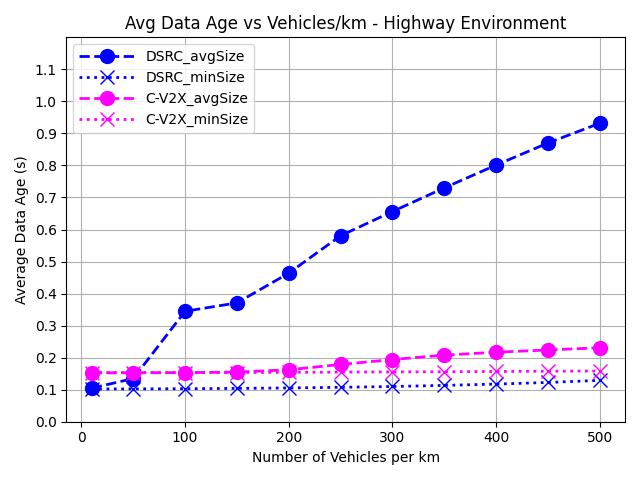}
  \label{fig:Data_age_highway}}
\subfloat[]{
  \includegraphics[width=0.5\columnwidth]{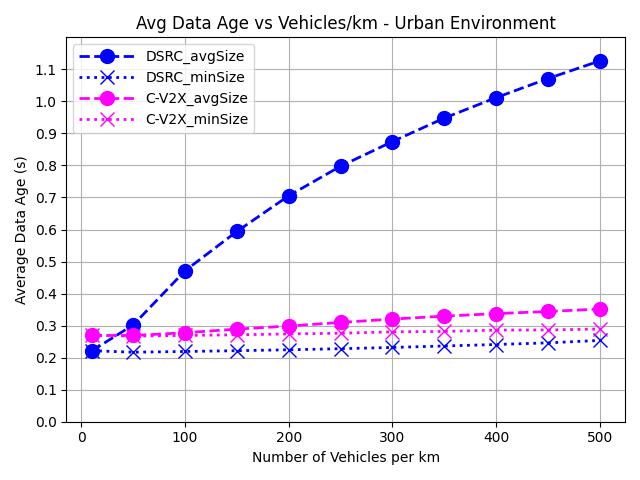}
  \label{fig:Data_age_urban}}
  \vfill
  \subfloat[]{  
\includegraphics[width=0.5\columnwidth]{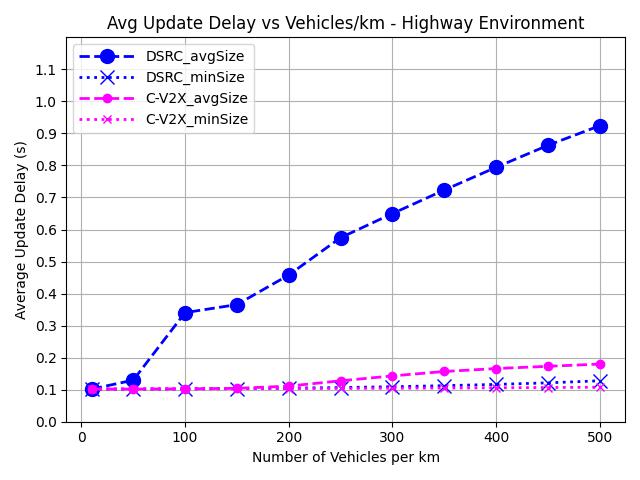}
  \label{fig:Update_delay_highway}}
\subfloat[]{
  \includegraphics[width=0.5\columnwidth]{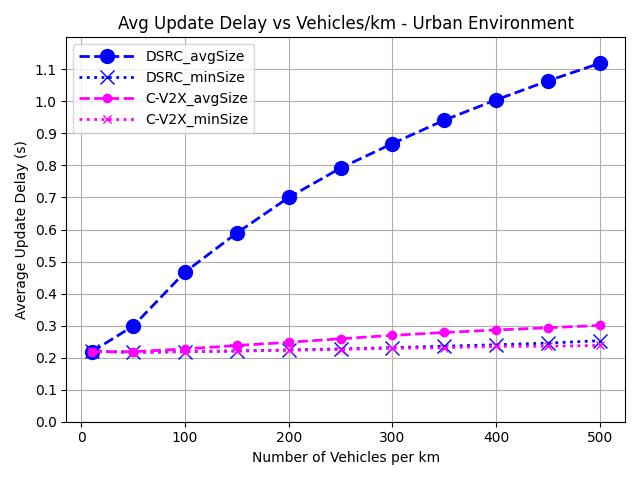}
  \label{fig:Update_delay_urban}}
\end{center} 
\caption{ (a) The Highway Data Age and (b) The Urban Data Age vs. the number of vehicles per km  (c) The Highway Update Delay and (d) The Urban Update Delay vs. the number of vehicles per km} 
\label{fig:delay}
\end{figure*}

From Figures \ref{fig:Data_age_highway}, \ref{fig:Data_age_urban}, we can find that increasing the number of vehicles/km increases the data age for both scenarios. The graphs show that the data age is higher in the urban scenarios than in the highway scenarios. This is due to that fact that urban environments are more congested with packets. When using DSRC RAT in our D-IM, the data age is much higher because of DSRC RAT use of CSMA/CA for madia access which cause longer latencies. Moreover, we can observe that the overhead of our implemented protocol leads to an increase of less than 0.032 seconds in data age for LTE-V2X simulations and less than 0.5 sec in DSRC simulations. The same can be concluded for the inter-packet gap (update delay) in Figures \ref{fig:Update_delay_highway}, \ref{fig:Update_delay_urban}. 
\begin{figure*}[ht]
\begin{center}
\subfloat[]{
  \includegraphics[width=0.5\columnwidth]{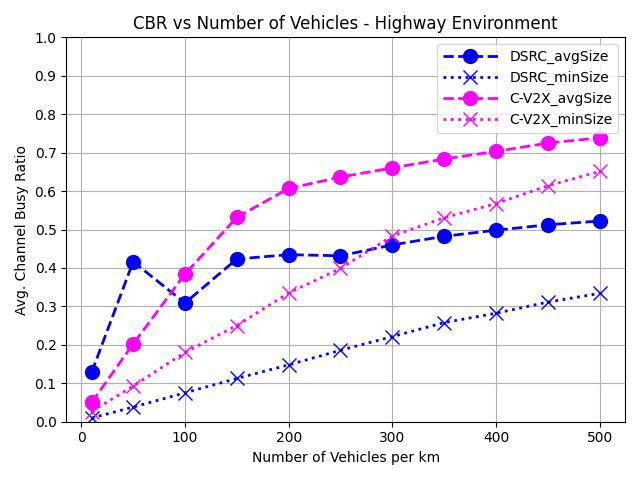}
  \label{fig:CBR_vehicles_highway}}
  \subfloat[]{
  \includegraphics[width=0.5\columnwidth]{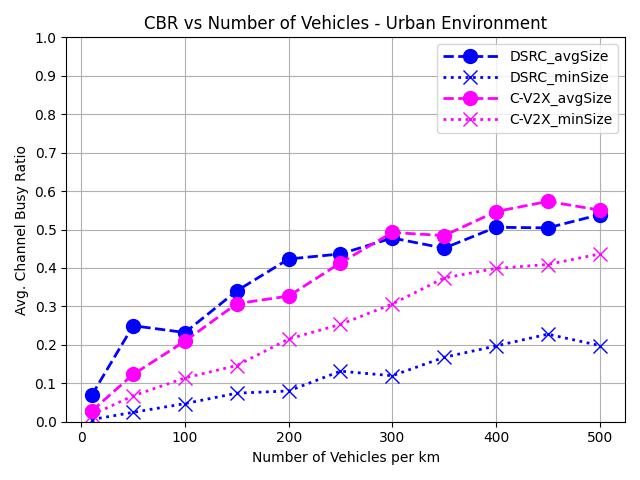}
  \label{fig:CBR_vehicles_urban}}
\end{center} 
\caption{ (a) Highway Channel Busy ratio and (b) Urban Channel Busy ratio vs. the number of vehicles per km} 
\label{fig:CBR_vehicles}
\end{figure*}

Figure \ref{fig:CBR_vehicles} shows the relationship between the CBR and the number of vehicles per km for both highway and urban scenarios. We can observe that increasing the number of vehicles/km increases the CBR for both environments. We also observe that the CBR is higher in the urban scenarios than in the highway scenarios when DSRC RAT is used. This is due to that fact that urban environments tend to have more vehicles than highway environments, resulting in more traffic and higher CBR given the fact that DSRC RAT uses "listen before talk"  strategy . The overhead of our implemented protocol leads to an average increase of less than 16\% in CBR when LTE-V2X is used, and less than 27\% when DSRC is used.

\begin{figure*}[t]
\begin{center}
\subfloat[]{  
\includegraphics[width=0.5\columnwidth]{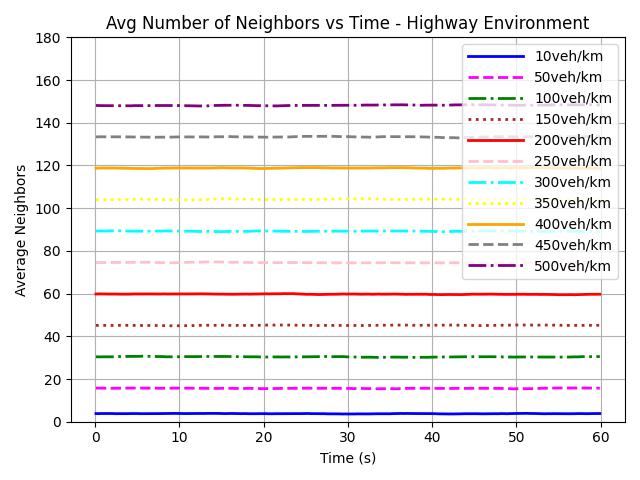}
  \label{fig:neighbors_highway}}
\subfloat[]{
  \includegraphics[width=0.5\columnwidth]{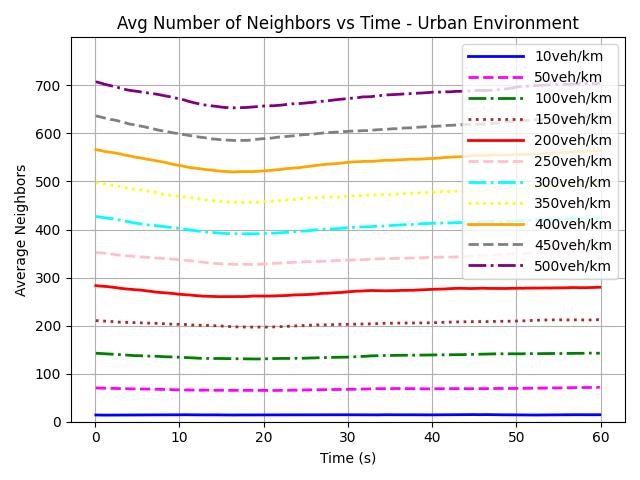}
  \label{fig:neighbors_urban}}
\end{center} 
\caption{ The number of neighbors vs. the number of vehicles per km for (a) Highway Scenarios and  (b) Urban Scenarios.} 
\label{fig:neighbors}
\end{figure*}
Figure \ref{fig:neighbors} studies the relationship between the number of neighbors and the number of vehicles per km for both highway Scenarios and urban Scenarios. We find an expected increase of the number of neighbors as we increase the number of vehicles/km for all scenarios. We also can see that the number of neighbors is much higher for urban scenarios that for the highway scenarios due to the nature of traffic. 

In summary, our simulation results validate the practicality and efficiency of the Zero Trust D-IM protocol. The framework's overhead on network performance is limited and well within acceptable limits for safety-critical V2X applications. Our findings show that even in dense traffic, the solution maintains robust network metrics. Specifically, in urban scenarios, the overhead from our protocol led to a manageable PRR degradation of around 18\% with LTE-V2X and 27\% with DSRC. For highway scenarios, the PRR reduction was a practical 6\% for LTE-V2X and 4\% for DSRC. Furthermore, our implementation added a minimal increase in Channel Busy Ratio (CBR) of less than 16\% for LTE-V2X (12\% for urban scenarios). Most critically for real-time applications, the latency overhead was negligible, with an increase of less than 0.032 seconds for LTE-V2X and under 0.5 seconds for DSRC. The maximum overhead observed and reported in this paper is for 500 vehicle/km density which is much more than the real-world  expected average vehicle density. The overhead for lower vehicle densities is much lower. These results collectively demonstrate that our blockchain-based solution can deliver robust, continuous identity verification and security guarantees without compromising the efficiency or reliability of the underlying communication network, without relying on implicit trust.

\section{CONCLUSIONS AND FUTURE WORK} \label{sec:cons}

This paper presented D-IM, a Zero Trust-based, blockchain-enabled decentralized identity management and authentication framework for autonomous vehicle (AV) communication systems. Motivated by the limitations of perimeter-based security models and centralized certificate authorities, D-IM was designed to address a comprehensive attacker model that captures identity-based, data integrity, collusion, availability, and trust-related threats in highly dynamic V2X environments.

We first defined a set of explicit design goals aligned with Zero Trust principles, including continuous verification, mutual authentication, decentralization, privacy preservation, non-repudiation, and traceability. Building on these goals, we introduced the D-IM system architecture and protocol, which leverage a permissioned blockchain as a decentralized trust anchor for identity registration, and verification. By binding vehicle identities to cryptographic credentials stored immutably on the blockchain and enforcing identity verification prior to any data exchange, D-IM eliminates implicit trust assumptions and reduces reliance on centralized authorities.

To evaluate the security properties of the proposed framework, we conducted both qualitative and formal security analyses, including a BAN logic-based verification of the identification and authorization protocol. The analysis demonstrates that D-IM achieves mutual authentication, fresh session key establishment, and resistance to impersonation, replay, man-in-the-middle, repudiation, and single-point-of-failure attacks under the defined threat model. Attacks that cannot be fully prevented at the identity layer, such as data falsification or collusion, are effectively detected, attributed, and contained through decentralized verification and accountability mechanisms.

We further validated the practicality of D-IM through simulation-based performance evaluations in realistic urban and highway scenarios using both DSRC and C-V2X technologies. The results show that continuous identity verification and blockchain-assisted authentication introduce limited communication and latency overhead, even under dense traffic conditions. Specifically, the framework maintains acceptable packet reception ratios, channel utilization, and end-to-end latency, demonstrating that strong security and accountability guarantees can be achieved without compromising the performance requirements of real-world AV deployments.

Overall, D-IM establishes a robust security foundation for autonomous vehicle communication networks by tightly integrating Zero Trust principles with decentralized identity management. The framework successfully addresses the defined attacker model while remaining scalable and efficient, making it well suited for next-generation intelligent transportation systems.

Building upon the foundational work presented in this paper, future research can focus on developing a comprehensive, multi-layered security framework for connected and autonomous vehicles. One key direction is to explore the integration of an advanced intrusion detection system that uses machine learning and deep learning models to analyze the behavior of vehicular communication protocols. Such a system could continuously assess the trustworthiness of entities in real-time and detect subtle, multi-layered attacks that are difficult to identify with traditional methods.

Another important area is the design of a decentralized consensus mechanism. This would allow a network of vehicles to collaboratively evaluate and confirm misbehavior without relying on a centralized authority. The use of smart contracts in this process could be further enhanced by incorporating Large Language Models (LLMs) to translate complex, human-readable security policies into executable code. This would enable dynamic and adaptable policy enforcement in a more intuitive and flexible manner. Finally, the development of a dynamic security policy is a crucial next step. This policy would automatically adjust a node's communication privileges based on confirmed malicious behavior, providing a proactive and adaptive system that maintains the overall integrity and stability of the network.

Future work will further strengthen the proposed framework through the integration of smart contract-based enforcement mechanisms and anomaly-based machine learning techniques. These extensions aim to automate misbehavior penalties, enhance detection of subtle or coordinated attacks, and improve system adaptability in evolving threat environments. Additionally, future research will explore large-scale deployments, more diverse adversarial behaviors, and deeper integration with cooperative perception and decision-making systems.

\bibliography{ref}

@article{he2022survey,
  title={A survey on zero trust architecture: Challenges and future trends},
  author={He, Yuanhang and Huang, Daochao and Chen, Lei and Ni, Yi and Ma, Xiangjie},
  journal={Wireless Communications and Mobile Computing},
  volume={2022},
  year={2022},
  publisher={Hindawi}
}

@article{anderson2023zero,
  title={A Zero Trust Architecture for Connected and Autonomous Vehicles},
  author={Anderson, John and Huang, Qiqing and Cheng, Long and Hu, Hongxin},
  journal={IEEE Internet Computing},
  year={2023},
  publisher={IEEE}
}

@inproceedings{wylde2021zero,
  title={Zero trust: Never trust, always verify},
  author={Wylde, Allison},
  booktitle={2021 international conference on cyber situational awareness, data analytics and assessment (cybersa)},
  pages={1--4},
  year={2021},
  organization={IEEE}
}

@article{li2022future,
  title={Future industry internet of things with zero-trust security},
  author={Li, Shan and Iqbal, Muddesar and Saxena, Neetesh},
  journal={Information Systems Frontiers},
  pages={1--14},
  year={2022},
  publisher={Springer}
}

@article{miglani2023blockchain,
  title={A Blockchain Based Matching Game for Content Sharing in Content-Centric Vehicle-to-Grid Network Scenarios},
  author={Miglani, Arzoo and Kumar, Neeraj},
  journal={IEEE Transactions on Intelligent Transportation Systems},
  year={2023},
  publisher={IEEE}
}

@article{dwivedi2023design,
  title={Design of blockchain and ecc-based robust and efficient batch authentication protocol for vehicular ad-hoc networks},
  author={Dwivedi, Sanjeev Kumar and Amin, Ruhul and Vollala, Satyanarayana and Das, Ashok Kumar},
  journal={IEEE Transactions on Intelligent Transportation Systems},
  year={2023},
  publisher={IEEE}
}

@article{dwivedi2021blockchain,
  title={Blockchain-based secured IPFS-enable event storage technique with authentication protocol in VANET},
  author={Dwivedi, Sanjeev Kumar and Amin, Ruhul and Vollala, Satyanarayana},
  journal={IEEE/CAA Journal of Automatica Sinica},
  volume={8},
  number={12},
  pages={1913--1922},
  year={2021},
  publisher={IEEE}
}

@article{vangala2020blockchain,
  title={Blockchain-enabled certificate-based authentication for vehicle accident detection and notification in intelligent transportation systems},
  author={Vangala, Anusha and Bera, Basudeb and Saha, Sourav and Das, Ashok Kumar and Kumar, Neeraj and Park, Youngho},
  journal={IEEE Sensors Journal},
  volume={21},
  number={14},
  pages={15824--15838},
  year={2020},
  publisher={IEEE}
}

@article{bazzi2019survey,
  title={Survey and perspectives of vehicular Wi-Fi versus sidelink cellular-V2X in the 5G era},
  author={Bazzi, Alessandro and Cecchini, Giammarco and Menarini, Michele and Masini, Barbara M and Zanella, Alberto},
  journal={Future Internet},
  volume={11},
  number={6},
  pages={122},
  year={2019},
  publisher={MDPI}
}

@misc{Iroha,title= {Hyperledger Iroha}, howpublished={\url{https://www.hyperledger.org/projects/iroha}} , note = {Accessed: 2024-05-01}}

@misc{Iroha_docs,title= {Hyperledger Iroha Documentation}, 
howpublished={\url{https://iroha.readthedocs.io/en/develop}} , note = {Accessed: 2024-05-01}}

@techreport{singh2015critical,
  title={Critical reasons for crashes investigated in the national motor vehicle crash causation survey},
  author={Singh, Santokh},
  year={2015}
}

@article{national2020early,
  title={Early Estimates of Motor Vehicle Traffic Fatalities and Fatality Rate by Sub-Categories Through June 2020},
  author={National Highway Traffic Safety Administration and others},
  year={2020}
}

@inproceedings{hoppe2008security,
  title={Security threats to automotive CAN networks--practical examples and selected short-term countermeasures},
  author={Hoppe, Tobias and Kiltz, Stefan and Dittmann, Jana},
  booktitle={Computer Safety, Reliability, and Security: 27th International Conference, SAFECOMP 2008 Newcastle upon Tyne, UK, September 22-25, 2008 Proceedings 27},
  pages={235--248},
  year={2008},
  organization={Springer}
}

@article{miller2014survey,
  title={A survey of remote automotive attack surfaces},
  author={Miller, Charlie and Valasek, Chris},
  journal={black hat USA},
  volume={2014},
  pages={94},
  year={2014}
}

@article{miller2019lessons,
  title={Lessons learned from hacking a car},
  author={Miller, Charlie},
  journal={IEEE Design \& Test},
  volume={36},
  number={6},
  pages={7--9},
  year={2019},
  publisher={IEEE}
}

@inproceedings{checkoway2011comprehensive,
  title={Comprehensive experimental analyses of automotive attack surfaces},
  author={Checkoway, Stephen and McCoy, Damon and Kantor, Brian and Anderson, Danny and Shacham, Hovav and Savage, Stefan and Koscher, Karl and Czeskis, Alexei and Roesner, Franziska and Kohno, Tadayoshi},
  booktitle={20th USENIX security symposium (USENIX Security 11)},
  year={2011}
}

@article{miller2015remote,
  title={Remote exploitation of an unaltered passenger vehicle},
  author={Miller, Charlie and Valasek, Chris},
  journal={Black Hat USA},
  volume={2015},
  number={S 91},
  pages={1--91},
  year={2015}
}

@article{psiaki2016attackers,
  title={Attackers can spoof navigation signals without our knowledge. Here's how to fight back GPS lies},
  author={Psiaki, Mark L and Humphreys, Todd E and Stauffer, Brian},
  journal={IEEE Spectrum},
  volume={53},
  number={8},
  pages={26--53},
  year={2016},
  publisher={IEEE}
}

@article{bhatia2019autonomous,
  author    = {M. Bhatia and A. Verma and A. Rao},
  title     = {Autonomous Vehicles: Control Systems and Challenges},
  journal   = {Journal of Automotive Technology},
  volume    = {14},
  number    = {3},
  pages     = {115--130},
  year      = {2019},
  publisher = {Journal of Automotive Technology}
}

@article{ying2024v2xReview,
  author    = {Z. Ying},
  title     = {A literature review on V2X communications security},
  journal   = {IET Intelligent Transport Systems},
  year      = {2024},
  doi       = {10.1049/cmu2.12778},
  url       = {https://ietresearch.onlinelibrary.wiley.com/doi/10.1049/cmu2.12778}
}

@inproceedings{ndss2024Revocation,
  author    = {G. Scopelliti and others},
  title     = {Efficient and Timely Revocation of V2X Credentials},
  booktitle = {Network and Distributed System Security Symposium (NDSS)},
  year      = {2024},
  url       = {https://www.ndss-symposium.org/wp-content/uploads/2024-17-paper.pdf}
}

@article{yousseef2025autonomous,
  title={Autonomous Vehicle Security: Hybrid Threat Modeling Approach},
  author={Yousseef, Amal and Lin, Yu-Zheng and Satam, Shalaka and Latibari, Banafsheh Saber and Pacheco, Jesus and Salehi, Soheil and Hariri, Salim and Satam, Pratik},
  journal={IEEE Open Journal of Vehicular Technology},
  year={2025},
  publisher={IEEE}
}

@phdthesis{satam2022autonomous,
  title={Autonomous Vehicle Security Framework (AVSF)},
  author={Satam, Shalaka},
  year={2022},
  school={The University of Arizona}
}

@article{annabi2025survey,
  author    = {Malak Annabi and Abdelhafid Zeroual and Nadhir Messai},
  title     = {Towards Zero Trust Security in Connected Vehicles: A Comprehensive Survey},
  journal   = {arXiv preprint arXiv:2504.05485},
  year      = {2025},
  url       = {https://arxiv.org/abs/2504.05485}
}

@article{ying2024literature,
  title     = {A literature review on V2X communications security: Foundation, solutions, status, and future},
  author    = {Zuobin Ying and Kaichao Wang and Jinbo Xiong and Maode Ma},
  journal   = {IET Communications},
  volume    = {18},
  number    = {20},
  pages     = {1683--1715},
  year      = {2024},
  publisher = {Wiley Online Library},
  doi       = {10.1049/cmu2.12778},
  url       = {https://ietresearch.onlinelibrary.wiley.com/doi/10.1049/cmu2.12778}
}

@article{hussain2018autonomous,
  title={Autonomous cars: Research results, issues, and future challenges},
  author={Hussain, Rasheed and Zeadally, Sherali},
  journal={IEEE Communications Surveys \& Tutorials},
  volume={21},
  number={2},
  pages={1275--1313},
  year={2018},
  publisher={IEEE}
}

@article{amoozadeh2015security,
  title={Security vulnerabilities of connected vehicle streams and their impact on cooperative driving},
  author={Amoozadeh, Mani and Raghuramu, Arun and Chuah, Chen-Nee and Ghosal, Dipak and Zhang, H Michael and Rowe, Jeff and Levitt, Karl},
  journal={IEEE Communications Magazine},
  volume={53},
  number={6},
  pages={126--132},
  year={2015},
  publisher={IEEE}
}

@article{luo2019cyberattacks,
  title={Cyberattacks and countermeasures for intelligent and connected vehicles},
  author={Luo, Feng and Hou, Shuo},
  journal={SAE International Journal of Passenger Cars-Electronic and Electrical Systems},
  volume={12},
  number={07-12-01-0005},
  pages={55--66},
  year={2019}
}

@article{durlik2024cybersecurity,
  title={Cybersecurity in autonomous vehicles—are we ready for the challenge?},
  author={Durlik, Irmina and Miller, Tymoteusz and Kostecka, Ewelina and Zwierzewicz, Zenon and {\L}obodzi{\'n}ska, Adrianna},
  journal={Electronics},
  volume={13},
  number={13},
  pages={2654},
  year={2024},
  publisher={MDPI}
}

@article{kim2024survey,
  title={A survey on adversarial robustness of lidar-based machine learning perception in autonomous vehicles},
  author={Kim, Junae and Kaur, Amardeep},
  journal={arXiv preprint arXiv:2411.13778},
  year={2024}
}

@article{sedar2023comprehensive,
  title={A comprehensive survey of V2X cybersecurity mechanisms and future research paths},
  author={Sedar, Roshan and Kalalas, Charalampos and V{\'a}zquez-Gallego, Francisco and Alonso, Luis and Alonso-Zarate, Jesus},
  journal={IEEE Open Journal of the Communications Society},
  volume={4},
  pages={325--391},
  year={2023},
  publisher={IEEE}
}

@article{kifor2024automotive,
  title={Automotive cybersecurity: A Survey on frameworks, standards, and testing and monitoring technologies},
  author={Kifor, Claudiu Vasile and Popescu, Aurelian},
  journal={Sensors},
  volume={24},
  number={18},
  pages={6139},
  year={2024},
  publisher={MDPI}
}

@article{matos2024survey,
  title={A survey on sensor failures in autonomous vehicles: Challenges and solutions},
  author={Matos, Francisco and Bernardino, Jorge and Dur{\~a}es, Jo{\~a}o and Cunha, Jo{\~a}o},
  journal={Sensors},
  volume={24},
  number={16},
  pages={5108},
  year={2024},
  publisher={MDPI}
}

@inproceedings{cao2023you,
  title={You can't see me: Physical removal attacks on $\{$lidar-based$\}$ autonomous vehicles driving frameworks},
  author={Cao, Yulong and Bhupathiraju, S Hrushikesh and Naghavi, Pirouz and Sugawara, Takeshi and Mao, Z Morley and Rampazzi, Sara},
  booktitle={32nd USENIX security symposium (USENIX Security 23)},
  pages={2993--3010},
  year={2023}
}

@inproceedings{sato2025realism,
  title={On the realism of lidar spoofing attacks against autonomous driving vehicle at high speed and long distance},
  author={Sato, Takami and Suzuki, Ryo and Hayakawa, Yuki and Ikeda, Kazuma and Sako, Ozora and Nagata, Rokuto and Yoshida, Ryo and Chen, Qi Alfred and Yoshioka, Kentaro},
  booktitle={Proceedings of the Network and Distributed System Security Symposium (NDSS)},
  year={2025}
}

@article{chahe2023dynamic,
  title={Dynamic adversarial attacks on autonomous driving systems},
  author={Chahe, Amirhosein and Wang, Chenan and Jeyapratap, Abhishek and Xu, Kaidi and Zhou, Lifeng},
  journal={arXiv preprint arXiv:2312.06701},
  year={2023}
}

@inproceedings{pavlitska2023adversarial,
  title={Adversarial attacks on traffic sign recognition: A survey},
  author={Pavlitska, Svetlana and Lambing, Nico and Z{\"o}llner, J Marius},
  booktitle={2023 3rd International conference on electrical, computer, communications and mechatronics engineering (ICECCME)},
  pages={1--6},
  year={2023},
  organization={IEEE}
}

@article{al2024can,
  title={CAN-MIRGU: a comprehensive CAN bus attack dataset from moving vehicles for intrusion detection system evaluation.},
  author={AL-KADRI, MO},
  year={2024}
}

@article{lampe2024can,
  title={Can-train-and-test: A curated CAN dataset for automotive intrusion detection},
  author={Lampe, Brooke and Meng, Weizhi},
  journal={Computers \& Security},
  volume={140},
  pages={103777},
  year={2024},
  publisher={Elsevier}
}

@article{twardokus2023toward,
  title={Toward protecting 5g sidelink scheduling in c-v2x against intelligent dos attacks},
  author={Twardokus, Geoff and Rahbari, Hanif},
  journal={IEEE Transactions on Wireless Communications},
  volume={22},
  number={11},
  pages={7273--7286},
  year={2023},
  publisher={IEEE}
}

@article{arif2024clustered,
  title={Clustered jamming in U-V2X communications with 3D antenna beam-width fluctuations},
  author={Arif, Mohammad and Kim, Wooseong},
  journal={Computer Communications},
  volume={216},
  pages={209--228},
  year={2024},
  publisher={Elsevier}
}

@article{abrar2024gps,
  title={GPS-IDS: An Anomaly-based GPS Spoofing Attack Detection Framework for Autonomous Vehicles},
  author={Abrar, Murad Mehrab and Youssef, Amal and Islam, Raian and Satam, Shalaka and Latibari, Banafsheh Saber and Hariri, Salim and Shao, Sicong and Salehi, Soheil and Satam, Pratik},
  journal={arXiv preprint arXiv:2405.08359},
  year={2024}
}

@ARTICLE{7426684,
  author={},
  journal={IEEE Std 1609.2-2016 (Revision of IEEE Std 1609.2-2013)}, 
  title={IEEE Standard for Wireless Access in Vehicular Environments--Security Services for Applications and Management Messages}, 
  year={2016},
  volume={},
  number={},
  pages={1-240},
  doi={10.1109/IEEESTD.2016.7426684}}

@techreport{etsi:103097,
  author        = {{ETSI Technical Committee ITS}},
  title         = {{Intelligent Transport Systems (ITS); Security; Security header and certificate formats}},
  institution   = {{European Telecommunications Standards Institute (ETSI)}},
  type          = {{TS}},
  number        = {{103 097}},
  version       = {{1.4.1}},
  date          = {{2020-10}},
  url           = {https://www.etsi.org/deliver/etsi_ts/103000_103099/103097/01.04.01_60/ts_103097v010401p.pdf},
  note          = {{Profile of IEEE Std 1609.2TM}}
}

@article{petit2014pseudonym,
  title={Pseudonym schemes in vehicular networks: A survey},
  author={Petit, Jonathan and Schaub, Florian and Feiri, Michael and Kargl, Frank},
  journal={IEEE communications surveys \& tutorials},
  volume={17},
  number={1},
  pages={228--255},
  year={2014},
  publisher={IEEE}
}

@article{khodaei2015key,
  title={The key to intelligent transportation: Identity and credential management in vehicular communication systems},
  author={Khodaei, Mohammad and Papadimitratos, Panos},
  journal={IEEE Vehicular Technology Magazine},
  volume={10},
  number={4},
  pages={63--69},
  year={2015},
  publisher={IEEE}
}

@article{almarshoud2024security,
  title={Security, privacy, and decentralized trust management in VANETs: A review of current research and future directions},
  author={AlMarshoud, Mishri and Sabir Kiraz, Mehmet and H. Al-Bayatti, Ali},
  journal={ACM Computing Surveys},
  volume={56},
  number={10},
  pages={1--39},
  year={2024},
  publisher={ACM New York, NY}
}

@article{farsimadan2025review,
  title={A review on security challenges in V2X communications technology for VANETs},
  author={Farsimadan, Eslam and Moradi, Leila and Palmieri, Francesco},
  journal={IEEE Access},
  year={2025},
  publisher={IEEE}
}

@article{yoshizawa2023survey,
  title={A survey of security and privacy issues in v2x communication systems},
  author={Yoshizawa, Takahito and Singel{\'e}e, Dave and Muehlberg, Jan Tobias and Delbruel, Stephane and Taherkordi, Amir and Hughes, Danny and Preneel, Bart},
  journal={ACM Computing Surveys},
  volume={55},
  number={9},
  pages={1--36},
  year={2023},
  publisher={ACM New York, NY}
}

@techreport{nist800207,
  author       = {Rose, Scott and Borchert, Oliver and Mitchell, Stu and Connelly, Sean},
  title        = {Zero Trust Architecture},
  institution  = {National Institute of Standards and Technology},
  type         = {NIST Special Publication},
  number       = {800-207},
  year         = {2020},
  doi          = {10.6028/NIST.SP.800-207},
  url          = {https://doi.org/10.6028/NIST.SP.800-207}
}

@techreport{dod2022zt,
  author       = {{U.S. Department of Defense}},
  title        = {DoD Zero Trust Strategy},
  institution  = {United States Department of Defense},
  year         = {2022},
  month        = {November},
  url          = {https://dodcio.defense.gov/Portals/0/Documents/Library/(U)DoD-Zero-Trust-Strategy.pdf}
}

@article{dhar2021securing,
  title={Securing IoT devices using zero trust and blockchain},
  author={Dhar, Suparna and Bose, Indranil},
  journal={Journal of Organizational Computing and Electronic Commerce},
  volume={31},
  number={1},
  pages={18--34},
  year={2021},
  publisher={Taylor \& Francis}
}

@article{li2022zero,
  title={Zero trust in edge computing environment: a blockchain based practical scheme},
  author={Li, Dawei and Zhang, Enzhun and Lei, Ming and Song, Chunxiao},
  journal={Mathematical Biosciences and Engineering},
  volume={19},
  number={4},
  pages={4196--4216},
  year={2022}
}

@article{maksymyuk2020blockchain,
  title={Blockchain-empowered framework for decentralized network management in 6G},
  author={Maksymyuk, Taras and Gazda, Juraj and Volosin, Marcel and Bugar, Gabriel and Horvath, Denis and Klymash, Mykhailo and Dohler, Mischa},
  journal={IEEE Communications Magazine},
  volume={58},
  number={9},
  pages={86--92},
  year={2020},
  publisher={IEEE}
}

@article{hassan2023blockchain,
  title={Blockchain and zero-trust identity management system for smart cities and IoT networks},
  author={Hassan, Yewande Goodness and Collins, Anuoluwapo and Babatunde, Gideon Opeyemi and Alabi, Abidemi Adeleye and Mustapha, Sikirat Damilola},
  journal={International Journal of Multidisciplinary Research and Growth Evaluation},
  volume={4},
  number={1},
  pages={704--709},
  year={2023}
}

@article{alevizos2022augmenting,
  title={Augmenting zero trust architecture to endpoints using blockchain: A state-of-the-art review},
  author={Alevizos, Lampis and Ta, Vinh Thong and Hashem Eiza, Max},
  journal={Security and privacy},
  volume={5},
  number={1},
  pages={e191},
  year={2022},
  publisher={Wiley Online Library}
}

@article{vcuvcko2021decentralized,
  title={Decentralized and self-sovereign identity: Systematic mapping study},
  author={{\v{C}}u{\v{c}}ko, {\v{S}}pela and Turkanovi{\'c}, Muhamed},
  journal={IEEE access},
  volume={9},
  pages={139009--139027},
  year={2021},
  publisher={IEEE}
}

@article{dib2020decentralized,
  title={Decentralized identity systems: Architecture, challenges, solutions and future directions},
  author={Dib, Omar and Rababah, Baha},
  journal={Annals of Emerging Technologies in Computing (AETiC)},
  volume={4},
  number={5},
  pages={19--40},
  year={2020},
  publisher={International Association for Educators and Researchers (IAER)}
}

@article{adja2021blockchain,
  title={A blockchain-based certificate revocation management and status verification system},
  author={Adja, Yves Christian Elloh and Hammi, Badis and Serhrouchni, Ahmed and Zeadally, Sherali},
  journal={Computers \& Security},
  volume={104},
  pages={102209},
  year={2021},
  publisher={Elsevier}
}

@article{das2023secure,
  title={A secure blockchain-enabled vehicle identity management framework for intelligent transportation systems},
  author={Das, Debashis and Dasgupta, Kousik and Biswas, Utpal},
  journal={Computers and Electrical Engineering},
  volume={105},
  pages={108535},
  year={2023},
  publisher={Elsevier}
}

@inproceedings{george2020secure,
  title={Secure identity management framework for vehicular ad-hoc network using blockchain},
  author={George, Sonia Alice and Jaekel, Arunita and Saini, Ikjot},
  booktitle={2020 IEEE Symposium on Computers and Communications (ISCC)},
  pages={1--6},
  year={2020},
  organization={IEEE}
}

@article{stockburger2021blockchain,
  title={Blockchain-enabled decentralized identity management: The case of self-sovereign identity in public transportation},
  author={Stockburger, Lukas and Kokosioulis, Georgios and Mukkamala, Alivelu and Mukkamala, Raghava Rao and Avital, Michel},
  journal={Blockchain: Research and Applications},
  volume={2},
  number={2},
  pages={100014},
  year={2021},
  publisher={Elsevier}
}

@article{noh2020distributed,
  title={Distributed blockchain-based message authentication scheme for connected vehicles},
  author={Noh, Jaewon and Jeon, Sangil and Cho, Sunghyun},
  journal={Electronics},
  volume={9},
  number={1},
  pages={74},
  year={2020},
  publisher={MDPI}
}

@article{tan2019secure,
  title={Secure authentication and key management with blockchain in VANETs},
  author={Tan, Haowen and Chung, Ilyong},
  journal={IEEE access},
  volume={8},
  pages={2482--2498},
  year={2019},
  publisher={IEEE}
}

@inproceedings{theodouli2020towards,
  title={Towards a blockchain-based identity and trust management framework for the IoV ecosystem},
  author={Theodouli, Anastasia and Moschou, Konstantinos and Votis, Konstantinos and Tzovaras, Dimitrios and Lauinger, Jan and Steinhorst, Sebastian},
  booktitle={2020 Global Internet of Things Summit (GIoTS)},
  pages={1--6},
  year={2020},
  organization={IEEE}
}

@article{liu2022blockchain,
  title={A blockchain-based decentralized, fair and authenticated information sharing scheme in zero trust internet-of-things},
  author={Liu, Yizhi and Hao, Xiaohan and Ren, Wei and Xiong, Ruoting and Zhu, Tianqing and Choo, Kim-Kwang Raymond and Min, Geyong},
  journal={IEEE Transactions on Computers},
  volume={72},
  number={2},
  pages={501--512},
  year={2022},
  publisher={IEEE}
}

@article{agudo2020blockchain,
  title={A blockchain approach for decentralized V2X (D-V2X)},
  author={Agudo, Isaac and Montenegro-G{\'o}mez, Manuel and Lopez, Javier},
  journal={IEEE Transactions on Vehicular Technology},
  volume={70},
  number={5},
  pages={4001--4010},
  year={2020},
  publisher={IEEE}
}

@inproceedings{van2017blackchain,
  title={Blackchain: Scalability for resource-constrained accountable vehicle-to-x communication},
  author={van der Heijden, Rens W and Engelmann, Felix and M{\"o}dinger, David and Sch{\"o}nig, Franziska and Kargl, Frank},
  booktitle={Proceedings of the 1st Workshop on Scalable and Resilient Infrastructures for Distributed Ledgers},
  pages={1--5},
  year={2017}
}

@article{cocirlea2020blockchain,
  title={Blockchain in intelligent transportation systems},
  author={Coc{\^\i}rlea, Drago{\c{s}} and Dobre, Ciprian and H{\^\i}r{\c{t}}an, Liviu-Adrian and Purnichescu-Purtan, Raluca},
  journal={Electronics},
  volume={9},
  number={10},
  pages={1682},
  year={2020},
  publisher={MDPI}
}

@article{das2023blockchain,
  title={Blockchain for intelligent transportation systems: Applications, challenges, and opportunities},
  author={Das, Debashis and Banerjee, Sourav and Chatterjee, Pushpita and Ghosh, Uttam and Biswas, Utpal},
  journal={IEEE Internet of Things Journal},
  volume={10},
  number={21},
  pages={18961--18970},
  year={2023},
  publisher={IEEE}
}

@inproceedings{yuan2016towards,
  title={Towards blockchain-based intelligent transportation systems},
  author={Yuan, Yong and Wang, Fei-Yue},
  booktitle={2016 IEEE 19th international conference on intelligent transportation systems (ITSC)},
  pages={2663--2668},
  year={2016},
  organization={IEEE}
}

@article{jabbar2022blockchain,
  title={Blockchain technology for intelligent transportation systems: A systematic literature review},
  author={Jabbar, Rateb and Dhib, Eya and Said, Ahmed Ben and Krichen, Moez and Fetais, Noora and Zaidan, Esmat and Barkaoui, Kamel},
  journal={IEEE Access},
  volume={10},
  pages={20995--21031},
  year={2022},
  publisher={IEEE}
}

@article{mollah2020blockchain,
  title={Blockchain for the internet of vehicles towards intelligent transportation systems: A survey},
  author={Mollah, Muhammad Baqer and Zhao, Jun and Niyato, Dusit and Guan, Yong Liang and Yuen, Chau and Sun, Sumei and Lam, Kwok-Yan and Koh, Leong Hai},
  journal={IEEE Internet of Things Journal},
  volume={8},
  number={6},
  pages={4157--4185},
  year={2020},
  publisher={IEEE}
}

@article{inedjaren2021blockchain,
  title={Blockchain-based distributed management system for trust in VANET},
  author={Inedjaren, Youssef and Maachaoui, Mohamed and Zeddini, Besma and Barbot, Jean-Pierre},
  journal={Vehicular Communications},
  volume={30},
  pages={100350},
  year={2021},
  publisher={Elsevier}
}

@inproceedings{li2021blockchain,
  title={A blockchain-based cooperative perception in internet of vehicles},
  author={Li, Xinghao and Tan, Chenchen and Liu, Minghao and Luan, Tom H and Gao, Longxiang and Qu, Youyang},
  booktitle={2021 IEEE 94th Vehicular Technology Conference (VTC2021-Fall)},
  pages={1--6},
  year={2021},
  organization={IEEE}
}

@article{song2020blockchain,
  title={Blockchain-enabled Internet of Vehicles with cooperative positioning: A deep neural network approach},
  author={Song, Yanxing and Fu, Yuchuan and Yu, F Richard and Zhou, Li},
  journal={IEEE Internet of Things Journal},
  volume={7},
  number={4},
  pages={3485--3498},
  year={2020},
  publisher={IEEE}
}

@inproceedings{zhu2024blockchain,
  title={A Blockchain-Based Accident Forensics System for Smart Connected Vehicles},
  author={Zhu, Caihua and Hu, Juhao and Wu, Jing and Long, Chengnian and Si, Xueming},
  booktitle={Proceedings of the 2024 6th Blockchain and Internet of Things Conference},
  pages={128--136},
  year={2024}
}

@article{cebe2018block4forensic,
  title={Block4forensic: An integrated lightweight blockchain framework for forensics applications of connected vehicles},
  author={Cebe, Mumin and Erdin, Enes and Akkaya, Kemal and Aksu, Hidayet and Uluagac, Selcuk},
  journal={IEEE communications magazine},
  volume={56},
  number={10},
  pages={50--57},
  year={2018},
  publisher={IEEE}
}

@TechReport{UCAM-CL-TR-138,
  author =	 {Burrows, Michael and Abadi, Mart{\'\i}n and Needham, Roger},
  title = 	 {{Authentication: a practical study in belief and action}},
  year = 	 1988,
  month = 	 jun,
  url = 	 {https://www.cl.cam.ac.uk/techreports/UCAM-CL-TR-138.pdf},
  institution =  {University of Cambridge, Computer Laboratory},
  doi = 	 {10.48456/tr-138},
  number = 	 {UCAM-CL-TR-138}
}

@article{burrows1990logic,
  title={A logic of authentication},
  author={Burrows, Michael and Abadi, Martin and Needham, Roger},
  journal={ACM Transactions on Computer Systems (TOCS)},
  volume={8},
  number={1},
  pages={18--36},
  year={1990},
  publisher={ACM New York, NY, USA}
}

@article{abdel2022proxy,
  title={A proxy signature-based swarm drone authentication with leader selection in 5G networks},
  author={Abdel-Malek, Mai A and Akkaya, Kemal and Bhuyan, Arupjyoti and Ibrahim, Ahmed S},
  journal={IEEE Access},
  volume={10},
  pages={57485--57498},
  year={2022},
  publisher={IEEE}
}

@article{farrea2025zero,
  title={Zero trust-based authentication for Inter-Satellite Links in NextGen Low Earth Orbit networks},
  author={Farrea, Kerry Anne and Baig, Zubair and Doss, Robin and Liu, Dongxi},
  journal={Ad Hoc Networks},
  volume={174},
  pages={103817},
  year={2025},
  publisher={Elsevier}
}

@article{guo2024uava,
  title={UAVA: Unmanned aerial vehicle assisted vehicular authentication scheme in edge computing networks},
  author={Guo, Zhenyang and Cao, Jin and Wang, Xinyi and Zhang, Yinghui and Niu, Ben and Li, Hui},
  journal={IEEE Internet of Things Journal},
  volume={11},
  number={12},
  pages={22091--22106},
  year={2024},
  publisher={IEEE}
}

\end{document}